\documentclass[sigconf,table,letterpaper]{acmart}

\usepackage{booktabs}

\usepackage[utf8]{inputenc}

\usepackage[english]{babel}

\usepackage{hyperref}
\usepackage{appendix}

\usepackage{arydshln}
\newcommand*\dhline{\hdashline[.4pt/3pt]}
\newcommand{\dcline}[1]{\cdashline{#1}[.4pt/3pt]}

\usepackage{fontawesome}

\usepackage{makecell}
\usepackage{multirow}

\usepackage{float}
\usepackage{placeins}

\usepackage{tabularx}

\usepackage[showboxes,absolute,overlay]{textpos}

\renewcommand{\texttt}[1]{%
	\begingroup
	\ttfamily
	\begingroup\lccode`~=`/\lowercase{\endgroup\def~}{/\discretionary{}{}{}}%
	\begingroup\lccode`~=`[\lowercase{\endgroup\def~}{[\discretionary{}{}{}}%
	\begingroup\lccode`~=`.\lowercase{\endgroup\def~}{.\discretionary{}{}{}}%
	\catcode`/=\active\catcode`[=\active\catcode`.=\active
	\scantokens{#1\noexpand}%
	\endgroup
}

\usepackage{balance}

\usepackage{subcaption}
\graphicspath{{figure/}{figures/}}

\copyrightyear{2021}
\acmYear{2021}
\setcopyright{iw3c2w3}
\acmConference[WWW '21]{Proceedings of the Web Conference 2021}{April 19--23, 2021}{Ljubljana, Slovenia}
\acmBooktitle{Proceedings of the Web Conference 2021 (WWW '21), April 19--23, 2021, Ljubljana, Slovenia}
\acmPrice{}
\acmDOI{10.1145/3442381.3450033}
\acmISBN{978-1-4503-8312-7/21/04}

\usepackage{pifont}
\newcommand{\cmark}{\ding{51}}%
\newcommand{\xmark}{\ding{56}}%

\usepackage{wasysym}
\newcommand{\full}{\CIRCLE}%
\newcommand{\half}{\LEFTcircle}%
\newcommand{\none}{\Circle}%

\usepackage{xspace}
\newcommand{\etal}{\textit{et~al.}}
\newcommand{\eg}{\textit{e.g.,}~}
\newcommand{\ie}{\textit{i.e.,}~}

\newcommand{\etc}{\textit{etc.}~}
\newcommand{\one}{({\em i})\xspace}
\newcommand{\two}{({\em ii})\xspace}
\newcommand{\three}{({\em iii})\xspace}

\makeatletter
\renewcommand{\paragraph}[1]{\vspace*{0.03in}\noindent{\bf #1.}\hspace{0.25ex \@plus1ex \@minus.2ex}}
\makeatother

\begin{document}
\title[Measuring Namespaces, Web Certificates, and DNSSEC of AA]{Security of Alerting Authorities in the WWW:\\ Measuring Namespaces, DNSSEC, and Web PKI}

\author{Pouyan Fotouhi Tehrani}
\affiliation{%
  \institution{Weizenbaum~Inst~/~Fraunhofer~FOKUS}
  \city{Berlin}
  \country{Germany}
}
\email{pft@acm.org}

\author{Eric Osterweil}
\affiliation{%
  \institution{George Mason University}
  \city{Fairfax}
  \state{VA}
  \country{USA}
}
\email{eoster@gmu.edu}

\author{Jochen H. Schiller}
\affiliation{%
  \institution{Freie Universit\"at Berlin}
  \city{Berlin}
  \country{Germany}
}
\email{jochen.schiller@fu-berlin.de}

\author{Thomas C. Schmidt}
\affiliation{%
  \institution{HAW Hamburg}
  \city{Hamburg}
  \country{Germany}
}
\email{t.schmidt@haw-hamburg.de}

\author{Matthias W\"ahlisch}
\affiliation{%
  \institution{Freie Universit\"at Berlin}
  \city{Berlin}
  \country{Germany}
}
\email{m.waehlisch@fu-berlin.de}

\renewcommand{\shortauthors}{Tehrani et al.}

\begin{abstract}
During disasters, crisis, and emergencies the public relies on online services provided by official authorities to receive timely alerts, trustworthy information, and access to relief programs.
It is therefore crucial for the authorities to reduce risks when accessing their online services.
This includes catering to secure identification of service, secure resolution of name to network service, and content security and privacy as a minimum base for trustworthy communication.

In this paper, we take a first look at \textit{Alerting Authorities} (AA) in the US and investigate security measures related to trustworthy and secure communication.
We study the domain namespace structure, DNSSEC penetration, and web certificates.
We introduce an integrative threat model to better understand whether and how the online presence and services of AAs are harmed.
As an illustrative  example, we investigate 1,388 Alerting Authorities. We observe partial heightened security relative to the global Internet trends, yet find cause for concern as about 78\% of service providers fail to deploy measures of trustworthy service provision.
Our analysis shows two major shortcomings. 
First, how the DNS ecosystem is leveraged: about 50\% of organizations do not own their dedicated domain names and are dependent on others, 55\% opt for unrestricted-use namespaces, which simplifies phishing, and less than 4\% of unique AA domain names are secured by DNSSEC, which can lead to DNS poisoning and possibly to certificate misissuance.
Second, how Web PKI certificates are utilized: 15\% of all hosts provide none or invalid certificates, thus cannot cater to confidentiality and data integrity, 64\% of the hosts provide domain validation certification that lack any identity information, and shared certificates have gained on popularity, which leads to fate-sharing and can be a cause for instability.
\end{abstract}

%
%



\maketitle

\setlength{\TPHorizModule}{\paperwidth}
\setlength{\TPVertModule}{\paperheight}
\TPMargin{5pt}
\begin{textblock}{0.8}(0.1,0.02)
	\noindent
	\footnotesize
	If you cite this paper, please use the WWW reference:
	Pouyan Fotouhi Tehrani, Eric Osterweil, Jochen H. Schiller, Thomas C.
	Schmidt, and Matthias Wählisch. 2021. Security of Alerting Authorities in
	the WWW: Measuring Namespaces, DNSSEC, and Web PKI. In \emph{Proceedings
	of the Web Conference 2021 (WWW ’21)}.
	ACM, New York, NY, USA, 12 pages. \url{https://doi.org/10.1145/3442381.3450033}
\end{textblock}

\section{Introduction}
\label{sec:intro}
Online media have been proven to be an effective channel to communicate with the public.
An ever growing number of Americans prefer to get their news online~\cite{pew-us-news19}, social media is being used for public health announcements~\cite{Tursunbayeva2017}, and authorities provide public disaster education and services via Web portals~\cite{FEMA2004}---just to mention a few examples.
Communication of critical information such as emergency response~\cite[Chapter 3]{PAHO2009} and provisioning of critical services are no exception to this trend.
Research shows that in emergencies the public turns to official and authoritative sources especially when specific, precise, and trustworthy information is requested~\cite{Condon2014,Endsley2014,Chauhan2017}.
At the same time, evaluating the credibility and trustworthines of online service providers during an emergency or crisis poses a real challenge for users~\cite{Manoj2007}.
A recent example to illustrate such situations is the novel Coronavirus (SARS-CoV-2) pandemic and its outbreak in the US in early 2020: with government institutions and health authorities being perceived as the most (social media being the least) trustworthy sources of information by the public~\cite{axios-trust20,trustpoll-20}, alone in the first month of the outbreak, nearly half a billion visits were registered on websites of Centers for Diseases Control and Prevention (CDC) and the National Institutes of Health~(NIH)~\cite{usanlytics}. Similarly, the high amount of visits on state unemployment websites brought the operation of many of those sites to a halt~\cite{nbc-crash20}.
The high demand for COVID-related online services took place in parallel with an explosion of misinformation campaigns and fraudulent services.
Despite efforts from top tech companies~\cite{techcrunch-joint20} the overwhelming \textit{infodemic}~\cite{who-infodemic20} continued to grow and prevail~\cite{ox-misinfo20}.
This over-abundance of information posed serious challenges both to politics and public health, and the growing number of individuals and business relying on unemployment insurance and governmental relief programs led to a boom in online fraud~\cite{nullSL2020} with many falling prey to such schemes.

\begin{figure}
	\centering
	\includegraphics{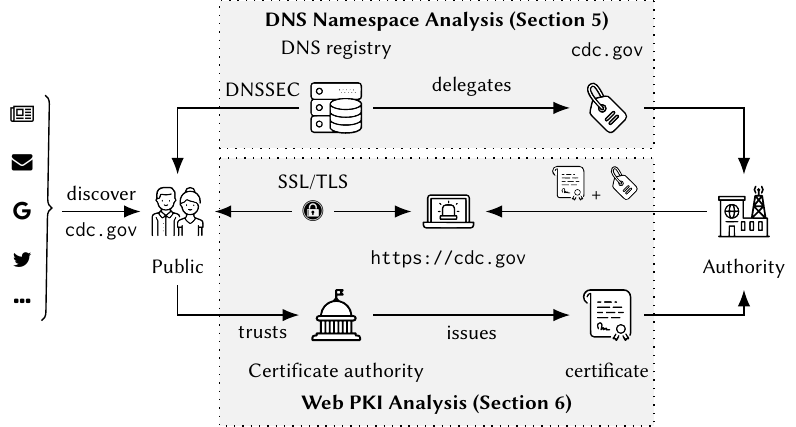}
	\caption{Accessing data from an Alerting Authority}
	\label{fig:overview}
\end{figure}

In this paper, we address the research blind spot of trustworthy and secure Web-based emergency services.
We systematically investigate the digital representation of emergency and disaster management organizations in the U.S. through the lens of the Domain Name System (DNS), its Security Extensions (DNSSEC)~\cite{RFC-4033,RFC-4034,RFC-4035}, \emph{and} the Web PKI (see \autoref{fig:overview}).
Based on our threat model, we aim  to understand whether and how specific integration of these organizations in the domain namespace and their use of DNSSEC and X.509 certificates can mitigate threats against trustworthy communication.
The point of departure for our study is the list of \textit{Alerting Authorities} (AA) provided by FEMA~\cite{fema-aa20}, which comprises all entities (the US governmental and non-governmental organizations) on federal, state, territorial, tribal, and local levels authorized to dispatch alerts.

Our key finding shows that only about 22\%, \ie 291 out of a total 1327 unique hosts, provide sufficient measures to ensure trustworthy identification.
This decomposes as follow:
\one only half the investigated organizations are uniquely identifiable based on dedicated domain names while the rise of multitenancy structures and shared certificates throughout the past decade has complicated identifications in general and has also led to an expansion of attack surfaces~\cite{osterweil2014shape}, 
\two the majority of organizations ($\approx 64\%$) do not take advantage of restricted namespaces for better protection against name spoofing and more than 96\% of investigated DNS zones are susceptible to DNS attacks due to lack of DNSSEC, and finally 
\three about 15\% are exposed to content poisoning as a result of invalid or no certificates.
In more detail, in this paper we contribute:

\begin{enumerate}
  \item \textbf{Threat model (Section~\ref{sec:trust})}.
  We introduce a threat model that integrates different characteristics of DNS and Web~PKI into groups of \emph{Assurances Profiles} that qualifies various degrees of reachable security.

  \item \textbf{Method (Section~\ref{sec:methodology})}.
  Our method identifies common public Alerting Authorities in the US and corresponding websites.
  The modular and configurable pipeline introduced here for data collection and analysis maintains a certain level of generality which makes it suitable to be extended to non-US regions in future work.

  \item \textbf{Analysis of namespace structure and protection~(Section~\ref{sec:dns-analysis})}.
  We map names of Alerting Authorities to fully qualified domain names (FQDN) and identify operational dependencies.
  Usage of restricted and protected namespaces as well as penetration of DNSSEC among AAs are investigated in this section.
	We also studied whether there are discrepancies between organizations from various fields of operation (\eg governmental, military).

  \item \textbf{Analysis of Web PKI~(Section~\ref{sec:x509-analysis})} We analyze Web PKI certificates used to authenticate and identify Alerting Authorities.
  On the one hand the historical and actual usage of X.509 are studied, and on the other hand it is investigated how widespread these technologies are, which certificate authorities are leading the market among AAs, and how (automated) domain-validation certificates affect trustworthy communication.
\end{enumerate}

While prior work has investigated the deployment of security protections broadly across different application domains, to the best of our knowledge, this is the first paper that investigates the security profiles of official critical (and critical-to-life) Alerting Authorities.
After presenting background and our results, we discuss improving measures and conclude with an outlook.

\section{Background}
\label{sec:bg}

Emergency management (EM) can be understood as an ongoing cycle of mitigating, preparing for, responding to, and recovering from incidents that threaten life, property, operations, or the environment~\cite{Blanchard2008,national2018nfpa}.
The core objectives of emergency management, ranging from coordination efforts to raising awareness and critical service provision, are carried out by governmental agencies, NGOs, volunteer groups, and international organizations.
The structure and organization of these entities differ in each country and even on local and regional levels.
In the US, the list of Alerting Authorities regularly published by FEMA~\cite{fema-aa20} provides a non-exhaustive overview of organizations which are (directly or indirectly) involved in the process of emergency management.

In each phase of EM cycle, communication (between and among authorities and the public) plays an integral role not just as a mere necessity but also in amounting to social resilience~\cite{Longstaff2008}.
Beside using dedicated alerting systems, \eg FEMA's \emph{Integrated Public Alert \& Warning System}~\cite{fema-ipaws20}, social media, or similar channels for information dissemination, many involved organizations have their own dedicated websites not only for informational purposes but also for services such as volunteer registry or disaster aid application~(\eg Homeland Security's \texttt{disasterassistance.gov}).

\begin{table*}
	\caption{Assurance profiles (\full~: strong, \half~: weak, \none~: inadequate) based on the interplay of DNS and X.509 certificate characteristics (\cmark: deployed, \xmark: not deployed) and security~implications for users (\faThumbsOUp~: fulfilled, \faWarning~: dependent protection, \faThumbsODown~: inadequate). Note if OV or EV certification is deployed, then domain validation is covered and not further assessed (--).}
	\label{tab:trust-matrix}
	\footnotesize
	\begin{minipage}{\textwidth}
		\setlength{\extrarowheight}{.25em}
		\begin{tabularx}{\textwidth}{l c c l c c l c c c X c}
			\toprule
			& \multicolumn{2}{c}{DNS} & & \multicolumn{2}{c}{Web PKI} & & \multicolumn{3}{c}{Security Implications} & & \multirow{2}{*}[-5pt]{\shortstack[c]{Assurance\\ Profile}}\\
			\cmidrule{2-3} \cmidrule{5-6} \cmidrule{8-10}
			\# & Restricted TLD & DNSSEC & & DV & OV/EV & & Identification & Resolution & Transaction & Weakness & \\
			\midrule
			01 & \cmark & \cmark & & -- & \cmark & & \faThumbsOUp & \faThumbsOUp & \faThumbsOUp & N/A & \full \\
			\midrule
			02 & \cmark & \cmark & & \cmark & \xmark & & \faWarning & \faThumbsOUp & \faThumbsOUp & Ambiguous identification & \half\\
			\dhline
			03 & \xmark & \cmark & & -- & \cmark & & \faWarning & \faThumbsOUp & \faThumbsOUp & Possible impersonation through name spoofing & \half\\
			\dhline
			04 & \cmark & \xmark & & -- & \cmark & & \faWarning & \faThumbsODown & \faThumbsOUp & DNS hijacking & \half\\
			\dhline
			05 & \xmark & \xmark & & -- & \cmark & & \faWarning & \faThumbsODown & \faThumbsOUp & Name spoofing, DNS hijacking & \half \\
			\midrule
			06 & \cmark & \xmark & & \cmark & \xmark & & \faWarning & \faThumbsODown & \faThumbsOUp & DNS hijacking and ambiguous identification & \none \\
			\dhline
			07 & \xmark & \xmark & & \cmark & \xmark & & \faThumbsODown & \faThumbsODown & \faThumbsOUp & Impersonation and DNS hijacking & \none \\
			\dhline
			08 & \xmark & \cmark & & \cmark & \xmark & & \faThumbsODown & \faThumbsOUp & \faThumbsOUp & Impersonation & \none \\
			\dhline
			09 & \cmark & \cmark & & \xmark & \xmark & & \faThumbsODown & \faThumbsOUp & \faThumbsODown & Content poisoning & \none \\
			\dhline
			10 & \cmark & \xmark & & \xmark & \xmark & & \faThumbsODown & \faThumbsODown & \faThumbsODown & DNS hijacking, content poisoning & \none \\
			\dhline
			11 & \xmark & \cmark & & \xmark & \xmark & & \faThumbsODown & \faThumbsOUp & \faThumbsODown & Impersonation, content poisoning & \none\\
			\dhline
			12 & \xmark & \xmark & & \xmark & \xmark & & \faThumbsODown & \faThumbsODown & \faThumbsODown & DNS hijacking, impersonation, content poisoning & \none\\
			\bottomrule
		\end{tabularx}
	\end{minipage}
\end{table*}

In this paper we investigate the namespace structure, DNSSEC penetration, and deployment of Web PKI certificates among Alerting Authorities to maintain secure communication (as defined in the next section).
The global domain name system (DNS), a distributed key-value database with a hierarchical namespace and management scheme, is de facto the entry point to many (if not all) of Internet services.
Respectively, for critical service providers, \eg Alerting Authorities, it is indispensable to be represented within namespaces protected both in organizational and technical terms: top-level domains (TLD) with restricted naming and delegation policies protect domain name owners against name and trademark violations while assuring end users that the domain name owner has undergone some form of vetting; at the same time, DNSSEC~\cite{RFC-4033} compensates the vulnerable client/server paradigm of DNS~\cite{RFC-3833} and caters for authenticated delegation and protect DNS data against tampering.
To authenticate the content provider behind a domain name X.509 certificates~\cite{RFC-5280} are used.
The semantics of a certificate depends on its certification process: if the real-world entity behind a certificate is vetted by a certification authority (CA) and is respectively awarded with an organization or extended validation certificate (OV/EV), the certificate can used for identification.
Otherwise, if the validation is limited to the ownership of a domain name, \ie domain validation (DV), the certificate is only good for authenticated confidentiality and integrity.

\section{A Threat Model for Web-Based Emergency Communication}
\label{sec:trust}
Emergency communication is dependent on heightened security requirements which are not always as relevant for other Internet services~(\eg video streaming, social media).
Three steps constitute our definition of secure online communication: \one~securely authenticating the authoritative service (``identification'' of the person, organization \etc behind the service name), \two~securely verifying 
that users have not been misdirected and are transacting with the service name they have identified (``resolution'' of name to network service), and \three~ensuring that the content was not altered, leaks privacy \etc during the session (``transaction'' security).
Although different methods can be utilized to realize such a secure workflow, here we focus on those technologies that are most accessible to (and deployable by) users and service operators in today's Internet, namely the DNS and the Web PKI ecosystems.
Alternative solutions are discussed in \autoref{sec:discus}.

\paragraph{An illustrative example}
For illustration of the communication workflow and respective security pitfalls, we consider the simple case of inquiring information about COVID-19 guidance as a resident of Jackson County in Missouri.
Through a search engine, an online ad, a recommendation from friends, \etc the URL is quickly discovered: \url{https://jacohd.org}.
When visiting the website, the presence of a green padlock in the address bar indicates \emph{only} the confidentiality and integrity of data exchange, but does not indicate whether the website belongs to the supposed service (\ie Health Department of Jackson County in Missouri instead of one of the other 22~Jackson Counties).
The generic domain name could haven been registered and operated by anyone.
An attacker could have published a forged website implementing the look and feel of the real health department.
At no stage is the user given the chance to authenticate the identity (\ie identification) of the service provider because the provided DV certificate does not include any identification information.
In contrast,
the health department of Jackson County in Michigan is reachable under \url{www.co.jackson.mi.us}\footnote{The complete URL is \url{https://www.co.jackson.mi.us/276/Health-Department}.}.
Here, the domain name under a restricted TLD indicates that it belongs to Jackson County (\texttt{co.jackson}) in Michigan~(\texttt{mi.us}), and the accompanying EV certificate serves as a definitive proof of identity.

\begin{figure*}
	\includegraphics{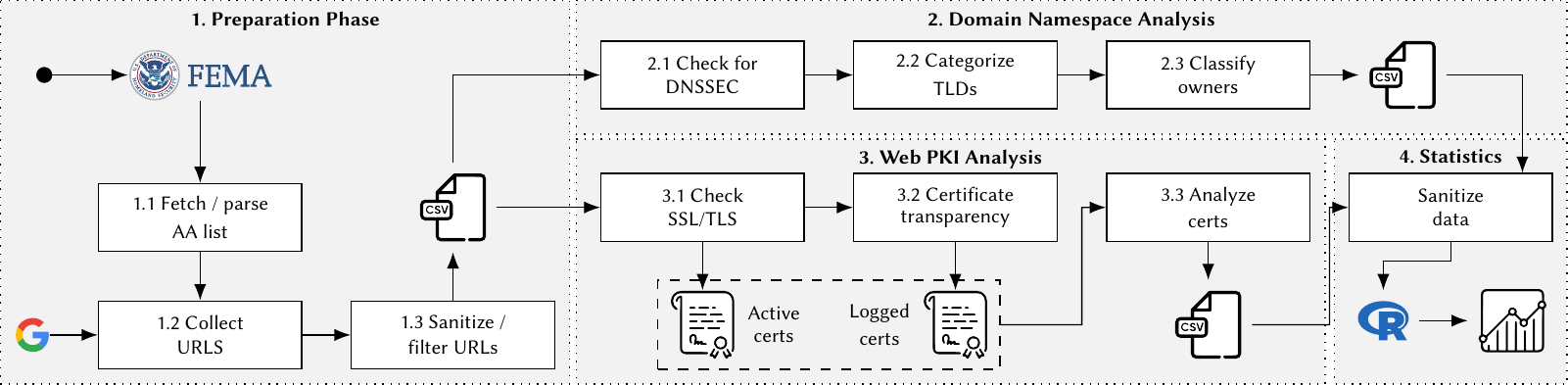}
	\caption{Toolchain to gather and analyze data about Alerting Authorities in the US}
	\label{fig:toolchain}
\end{figure*}

\paragraph{Threaten identification}
In a secure setting, it would be possible to identify and authenticate the communication partner \emph{before} initiating the transaction.
Yet, the point of departure for Web communication are domain names, which cannot be used for secure identification, while Web PKI certificates (as proofs of identity) are provided only \emph{after} resolution succeeds and transaction is initiated.
This implies that targeting authentic names and subsequent secure resolution are necessary (yet insufficient) conditions of identification through a certificate.
Respectively, simple name spoofing, \eg through typosquatting~\cite{khan2015every}, DNS cache poisoning~\cite{wang2015revisit}, or other DNS hijacking attacks, which can mislead users to malicious services, can act as a precursor for impersonation attacks, especially if subsequently only a DV certificate with no identity information is presented.
A viable countermeasure is the use of restricted namespaces so at least the affiliation or identity of the service provider can be inferred directly from its respective name.
Governmental organizations in the US, for example, educate visitors that domain names of federal government agencies most commonly end in \texttt{.gov} or \texttt{.mil}.
Subsequently, an OV/EV certificate provides direct elements of proof of identity.
When considering using approaches like this, exceptions may serve to help prove the rule: consider that the United States Post Office's (USPO's) official website is {\tt uspo.{\bf com}}, \ie not under {\tt .gov}.
This, then, necessitates additional knowledge or verification before users of \emph{that} government agency can be assured that they are transacting with the official authority online.
Proper identification, thus, involves both selection of proper domain names, secure resolution (see below), and identity information from OV/EV certificate.

\paragraph{Threaten resolution}
The second attack surface pertains to name resolution.
There are two methods to assert that a name has been resolved correctly: either by using DNSSEC or through the X.509 bindings in a certificate's common name (CN) or subject alternative names (SAN).
The latter approach, however, provides only an a posteriori assurance, \ie after transaction initiation with a server, and only if the resolution has already succeeded correctly.
Furthermore, especially in case of DV certificates, if resolution is compromised, \eg through DNS poisoning, and as a result attackers were granted a DV certificate~(see Brandt~\emph{et~al.}~\cite{Brandt2018}), there is no way to verify the integrity of the resolution process.
The only effective solution in securing name resolution and deterring collateral damages such as DV certificate misissuances is deployment of DNSSEC.

\paragraph{Threaten transaction}
In the final step, after name identification and resolution, it is imperative to secure the transaction using transport security protocols (TLS/SSL) in terms of authenticated confidentiality and integrity.
It is worth noting that authentication (using X.509 certificates) is crucial, because encryption and integrity checks alone can also be performed by a malicious actor using monkey-in-the-middle attacks.

\paragraph{Assurance profiles}
Based on threats on the three aforementioned dimensions, \autoref{tab:trust-matrix} provides the various combinations of DNS and Web PKI options and the security implications of their deployment for users; the \emph{Assurance Profiles} summarizes their combinations.

To achieve \emph{strong assurance}, a service provider should own a domain name that \one indicates its affiliation, \two is securely delegated, and \three is bound with a real-world entity through an OV/EV certificate.
A prominent example is \url{coronavirus.gov} which is registered under \texttt{.gov} TLD denoting it being a governmental domain name, supports DNSSEC (\ie cannot be hijacked), and provides a valid OV certificate belonging to the \emph{Executive Office of the President}.

A service provider that only partially covers these aspects and fails to deploy DNSSEC or uses a name under a non-restricted namespace exhibits a \emph{weak Assurance Profile} due to susceptibility to DNS hijacking or simple name spoofing.
This is how a campaign in Germany was able to defraud up to 4000 applicants of the corona relief program of a federal state.
The scammers spoofed the original domain name \url{soforthilfe-corona.nrw.de} by registering \url{nrw-corona-soforthilfe.de} without much burden because the \texttt{.de} TLD has no delegation restrictions that cannot be circumvented with minimal effort.
Although the authentic name (the former) was bound to a valid OV certificate, the spoofed name was awarded with a DV certificate which gave the impression of authenticity and caused users to fell prey to this phishing campaign.
Similarly, the threat of DNS hijacking was highlighted during the pandemic as attackers managed to exploit a vulnerability in home routers and made use of insecure DNS to manipulate name resolution; an attack which could easily be defended through DNSSEC.

In contrast to the previous cases, \emph{inadequate assurance} reflects the case when no certificate or only a DV certificate is provided \emph{regardless} of domain name properties and presence of DNSSEC.
Lack of a certificate at the very least defeats the purpose of authenticated encryption and integrity verification\footnote{Considering that alternative SSL/TLS authentication methods, \eg pre-shared keys, are not scalable and suitable for studied cases here.}, while a mere DV certificate can at best only cater to confidentiality and integrity without providing any information about the identity of service provider.
An example of weak assurance is the Corona Emergency Response Fund of CDC foundation under \url{give4cdcf.org} which have raised millions of dollars in fighting the pandemic.
The usage of \texttt{.org} generic TLD simplifies name spoofing\footnote{At the time of writing \url{give4cdcf.net} remains undelegated.}, lack of DNSSEC make it a suitable target for hijacking, and finally the provided DV certificate practically doesn't provide any evidence of identity.

\begin{table*}
	\caption{Top-level domains in use by Alerting Authorities}
	\footnotesize
	\begin{minipage}{\textwidth}
		\setlength{\extrarowheight}{.25em}
		\begin{tabularx}{\textwidth}{l l c l c r l X c l r r r}
			\toprule
			& \multicolumn{2}{c}{TLD} & & \multicolumn{2}{c}{Registration} & & \multicolumn{2}{c}{Registry} & & \multicolumn{3}{c}{Statistics} \\
			\cline{2-3} \cline{5-6} \cline {8-9} \cline{11-13}
			Type & Label & DNSSEC & & Restricted & {\centering Fee/year} & & Name & Country & & Share & Count & DNSSEC \\
			\hline
			\multirow{5}{*}{gTLD} & \texttt{.com} & \cmark & & \xmark & < 15~\$ & & Verisign & US & & 19.44 \% & 258 & 2\\
			\dcline{2-10}
			& \texttt{.org} & \cmark & & \xmark & < 15~\$ & & Public Internet Registry & US & & 26.15 \% & 347 & 5\\
			\dcline{2-10}
			& \texttt{.net} & \cmark & & \xmark & < 15~\$ & & Verisign & US & & 4.37 \% & 58 & 0\\
			\dcline{2-10}
			& \texttt{.info} & \cmark & & \xmark & < 15~\$ & & Afilias & US & & 0.15 \% & 2 & 0\\
			\dcline{2-10}\cline{11-13}
			\multicolumn{10}{r}{~} & 50 \% & 665 & 7\\
			\hline
			\multirow{4}{*}{ccTLD} & \texttt{.cc} & \cmark & & \xmark & < 15~\$ & & eNIC~\textsuperscript{1} & US & & 0.07 \% & 1 & 0\\
			\dcline{2-10}
			& \texttt{.co} & \cmark & & \xmark & < 20~\$ & & .CO Internet S.A.S~\textsuperscript{2} & US & & 0.07 \% & 1 & 0\\
			\dcline{2-10}
			& \texttt{.us} & \cmark & & (\cmark) & < 15~\$ & & Neustar & US & & 4.89 \% & 65 & 0\\
			\dcline{2-10}\cline{11-13}
			\multicolumn{10}{r}{~} & 5.04 \% & 67 & 0\\
			\hline
			ccSLD & \texttt{.<code>.us} & (\cmark) & & \cmark & -- & & Neustar & US & & 17.71 \% & 235 & 2\\
			\hline
			\multirow{4}{*}{sTLD} & \texttt{.edu} & \cmark & & \cmark & 77~\$ & & Educase~\textsuperscript{3} & US & & 0.45 \% & 6 & 0\\
			\dcline{2-10}
			& \texttt{.gov} & \cmark & & \cmark & 400~\$ & & General Services Administration & US & & 25.92 \% & 344 & 30\\
			\dcline{2-10}
			& \texttt{.mil} & \cmark & & \cmark & -- & & Defense Information Systems Agency & US & & 0.75 \% & 10 & 10\\
			\dcline{2-10}\cline{11-13}
			\multicolumn{10}{r}{~} & 27.12 \% & 360 & 40\\
			\hline
			\multicolumn{11}{r}{Unique domain names} & 1327 & \\
			\bottomrule
		\end{tabularx}
		\smallskip
		
		{\scriptsize\textsuperscript{1} subsidiary of Verisign, \textsuperscript{2} subsidiary of Neustar, \textsuperscript{3} operated by Verisign}
		\label{tab:domain-type-freq}
	\end{minipage}
\end{table*}

\section{Method and Data Corpus}
\label{sec:methodology}
The subject of study in this paper are the US organizations involved in EM.
Due to lack of a central registry, we focus on the list of Alerting Authorities maintained by FEMA. 
Although this list might not include each and every entity involved in emergency management, it provides a decent, legitimate overview over this field comprising a wide spectrum of organizations ranging from local governments, law enforcement agencies, and military bases to NGOs and universities.
Each entry represents an organization by a unique ID, a name, and a territory of operation (including unincorporated territories).
Throughout this study, we use the AA~list from September~11,~2019 comprising 1,388 entries (excluding a single duplicate entry). 

Our method consists of three~phases: (1) preparation phase, (2) domain namespace analysis, and (3) Web PKI analysis.
Our measurements were carried out from October~2019 up to March~2020 with each measurement being executed at least twice from various vantage points in Europe and the US to detect any possible vantage point dependent discrepancies, \eg limited access due to geo-blocking.
\autoref{fig:toolchain} summarizes our methodology from preparation phase to data gathering and final analysis (see \S~\ref{sec:dns-analysis} and \S~\ref{sec:x509-analysis}).

\paragraph{(1) Preparation}
In the preparation phase, we first retrieve and parse the AA list and assign the domain name used for web services for each organization. 
To identify the primary website of an Alerting Authority, we query and scrape the Google search engine.
For each entry in the AA list, the combination of name and territory of operation (\eg \textit{Fresno Police Department CA}) was used as query string.
Each query yielded between 4 and 12 results.
Since the results are not necessarily ranked to have the official URL first, we excluded results based on a list of inapt domain names (\eg social media sites and yellow pages).
The topmost remaining URL was then selected for the respective organization.
Finally, the list of collected URLs was manually checked to remove any mismatches and falsely associated URLs which were not detected automatically, \eg same URL for homonymous counties in different states.
A total of $23$ entries were removed: $11$ entries with mismatched names, $11$ associated with the wrong territory of operation, and $1$ with no matching URL at all; leaving a total of $1,365$ URLs for further analysis.
The remaining URLs (\eg \url{https://www.fresno.gov/police}) were parsed to extract the FQDNs (\eg \texttt{www.fresno.gov}) and path segments (\eg \texttt{/police}).

\paragraph{(2) Domain Namespace Analysis}
In the second phase we first separate effective second-level domains (SLD) from TLDs, \eg for `\url{www.ci.tracy.ca.us}', `\texttt{ca.us}' being the TLD (more specifically the public suffix) and `\texttt{tracy}' the effective SLD.
We then check DNSSEC status for both the given domain name and its TLD, categorize TLDs (restricted/unrestricted), and finally, based on a list of predefined keywords (see \autoref{tab:domain-classes} in \hyperref[sec:app]{Appendix}) map each Alerting Authority to a \emph{field of activity} as either Public safety, Governmental, Law enforcement, Military, or Educational.
The results of our analysis on domains names is presented in Section~\ref{sec:dns-analysis}.

\paragraph{(3) Web PKI Analysis}
Finally, domain names were used to investigate the current and historic adaption of Web PKI certificates 
by respective hosts.
To study the current state, OpenSSL version \texttt{1.1.1d} CLI was leveraged to fetch complete certificate chains, perform validation, and verify revocation status using stapled Online Certificate Status Protocol (OCSP)~\cite{RFC-6066}, manual OCSP~\cite{RFC-6960}, or Certificate Revocation Lists (CRL)~\cite{RFC-5280} (depending on availability).
For our historical analysis, we used CT logs~\cite{RFC-6962,Scheitle2018}.
To do this we leveraged the publicly accessible database provided by \textit{Sectigo} under \href{https://crt.sh/}{crt.sh}, which audits 79 log servers from 12 organizations (at the time of writing).
For any given host name, the database was queried for certificates which have the host name or a wildcard covering the host name as their subject name or have it included in the list of subject alternative names (SAN).
From a total of $28,370$ retrieved unique certificates, $10,826$ were pre-certificates and are omitted from further analysis.
The remaining $17,544$ certificates were then limited to those issued in the past decade (2009-2019), leaving a total number of $17,477$ certificates which are analyzed as described in~Section~\ref{sec:x509-analysis}.

\section{DNS Namespace Analysis}
\label{sec:dns-analysis}
By studying the domain names of alerting authorities, we aim to answer the following questions:

\begin{enumerate}
	\item Does each AA have its own dedicated domain name?
	\item How do AAs integrate in the global DNS namespace?
	\item Do AAs secure their names using DNSSEC?
\end{enumerate}

\noindent
The first question is concerned with how Alerting Authorities maintain their online presence, and avoid unnecessary dependencies.
Lack of a dedicated name, for example, leads to dependence on someone else for authentication and data security as X.509 certificates are bound to domain names.
The second question aims to investigate whether AAs prefer specific TLDs to take advantages of recognizability (\eg governmental organization under \texttt{.gov}) and security (restricted vs. non-restricted TLDs).
Finally, the last question regards measures taken in securing names against threats such as spoofing or DNS hijacking which can also lead to impersonation and phishing.
\autoref{tab:domain-type-freq} summarizes our findings.

\subsection{Dedicated Domain Names}
We consider an AA to have a dedicated DNS name either if it has its own effective SLD, or has been assigned a sub-domain under the namespace of its parenting organization or any generic service provider, which is not shared.
For example, the \textit{Tehama County Sheriff} (\href{https://tehamaso.org/}{\texttt{tehamaso.org}}) has its own dedicated name whereas \textit{Apache County Sheriffs Office} (\url{www.co.apache.az.us/sheriff/}) does not.

To measure dedicated domain names we divided the set of AA URLs into two groups depending on whether the URL path segment is empty (674 entries) or not (691 entries); the group with empty path segments was then regarded as having dedicated names.
To prevent false positives of non-dedicated names, we manually examined all these websites and verified that the landing page does not relate to the Alerting Authority.
We found only 25~false positives (\eg \url{http://www.franklincountyema.org/db/} with \texttt{/db} path being the start page), which leads to overall $\approx$~51\% AAs with dedicated names while the rest represents common names of parent organizations or other service providers.
We also observed three emergency management agencies with dedicated names which are redirected (using HTTP 301/302 response codes) to web pages under county or state websites.
Out of the total 1,365~collected URLs 1,327 unique domain names exist, showing that in some cases multiple entities are subsumed under the same domain, \eg different agencies all under the domain name of a single state.

The data also shows that all educational entities (total of 4) and over 90\% of governmental entities (467 out of 503) such as state and local governments own dedicated names in contrast to only $\approx$~25\% of public safety entities (164 out of 669), and less than half of military organizations (8 out of 19) which nearly all are represented under \href{https://home.army.mil}{\texttt{home.army.mil}}.

\subsection{Namespace Structure}
We start with various TLDs and country code second-level domains (ccSLDs) in use by Alerting Authorities, which we group as follows:

\begin{description}
	\item[gTLD~\cite{icann-gtld}:] generic top-level domain, \eg \texttt{.org}
	\item[ccTLD~\cite{icann-cctld}:] country code top-level domain, \eg \texttt{.us}
	\item[ccSLD~\cite{icann-sld}:] country code second-level domain, \eg \texttt{.ny.us}
	\item[sTLD~\cite{icann-stld}:] sponsored top-level domain, \eg \texttt{.mil}
\end{description}

\noindent
Each TLD group features different properties.
In general, there are little to no delegation limits and naming conventions for names under gTLDs or ccTLDs except for the \texttt{.us} namespace.
Under \texttt{.us} ccTLD more than 3,000~names are reserved and unavailable for public registration~\cite{dotus-faqs} and the namespace has a rigorous structure with domain names at second, third, or fourth levels.
This structuring reflects the ``political geography''~\cite{RFC-1480} and defines a number of reserved names for designated organizations or purposes, \eg county or city, and territory of operation~\cite{RFC-1480,uscompliance}. 
Finally, sponsored TLDs  (\texttt{.edu}, \texttt{.gov}, and \texttt{.mil}) impose stricter eligibility requirements and thus have an advantage over gTLD names so that it can be made sure that only eligible registrants are granted the ownership of respective domain names~\cite{RFC-920,RFC-2146}, given that such policies are adequately enforced by respective registries.

As summarized in \autoref{tab:domain-type-freq}, whereas half of domain names are registered under generic TLDs, the remaining majority~($\approx$~45\%) makes use of sponsored TLDs and names within the \texttt{.us} state-code namespace, and the rest 5 percent opts for domains under ccTLDs.
It is noteworthy that the \texttt{.us} locality namespace exhibits a relatively low penetration among AAs.
For example, the usage of canonical forms \texttt{[\textbf{ci},\textbf{co}].<locality>.<state-code>.us} for cities or counties: we observe that for every 5 cities which have the term \textit{city} in their domain names there exists only 1 city which uses the foreseen  naming pattern, and for every 4 counties choosing to have the term \textit{county} in its domain name, there is only one county opting for the canonical form.

\begin{figure}
	\centering
	\footnotesize
	\includegraphics{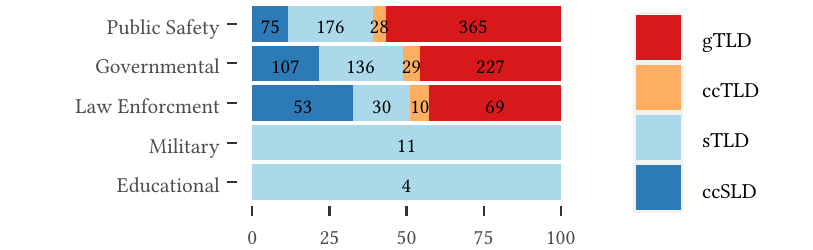}
	\caption{Distribution of TLD types per operation territory}
	\label{fig:domain-type-per-org-barplot}
\end{figure}

Finally, we examined if the specific choice of top-level domains for an organization correlates with the organization's field of operation.
\autoref{fig:domain-type-per-org-barplot} depicts how widespread various TLD types are in use in different fields of operation. 
It is noteworthy that educational and military organizations make exclusive use of restricted TLDs (\texttt{.edu} and \texttt{.mil} respectively), whereas gTLDs remain the more popular choice among the others.
This figure also confirms the previous observations that the majority of remaining organizations, regardless of field of operation, opt for generic TLDs instead of taking advantage of special namespaces within the well-organized structured of \texttt{.us} namespace.

\subsection{DNSSEC Deployment}
\label{subsec:dnssec}
\begin{figure}[b]
	\centering
	\footnotesize
	\includegraphics{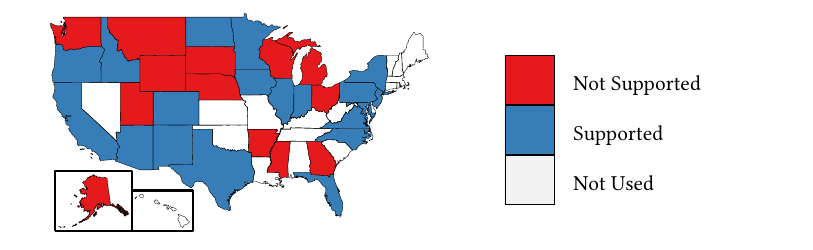}
	\caption{Support for DNSSEC among \texttt{.us} ccSLDs in use}
	\label{fig:dnssec-per-state-map}
\end{figure}

We used \texttt{drill} to chase DNS signatures and verify if a domain has properly activated DNSSEC.
All TLDs in use by AAs (see \autoref{tab:domain-type-freq}) support DNSSEC except a number of \texttt{.us} ccSLD domains: out of 50 total state ccSLDs under \texttt{.us} namespace, 32 have been used by AA organizations with only 18 supporting DNSSEC.
\autoref{fig:dnssec-per-state-map} depicts the state ccSLDs, which support DNSSEC (blue), which do not support (red), and those which are not used by any of organizations in our data set (white).

Although $\approx$~57\% of TLDs in use support DNSSEC, less than 4\% of AA~domain names have DNSSEC enabled.
Compared with the longitudinal DNSSEC study of Chung et al.~\cite{Chung2017}, measuring $0.6\%$ for \texttt{.com} and $1.0\%$ for \texttt{.org} domains, we observe a higher DNSSEC penetration.
However, to our surprise even among \texttt{.gov} SLDs which are mandated to implement DNSSEC~\cite{Technology2009} less than 10\% (30) have support for DNSSEC which is considerably less than the $\approx 90\%$ DNSSEC penetration among select governmental organizations (sample set of ca. 1200 \texttt{.gov} SLDs) as measured by NIST~\cite{dnssec-gov}.

\section{Web PKI Analysis}
\label{sec:x509-analysis}
The ecosystem of Web PKI revolves around X.509 certificates.
We investigate the deployment and characteristics of certificates in the context of Alerting Authorities to answer the following questions:

\begin{enumerate}
	\item To what extent do AAs adapt web PKI?
	\item How is the historic landscape of X.509 shaped among AAs?
\end{enumerate}

\subsection{Current Deployment of Certificates}
\label{subsec:webpkit-deployment}
To have a better understanding of the current deployment of web certificates, we gathered a snapshot of SSL/TLS deployment on public servers of Alerting Authorities.

Out of the total 1327 unique names, 1187 hosts ($\approx$~89\%) support SSL/TLS with 1130 hosts ($\approx$~95\%) delivering valid X.509 certificates.
Within the remaining 57 hosts, 17 use expired certificates, 9 use self-signed certificates, and 1 has self-signed certificates in its certificate chain.
The validity of certificates provided by the remaining 30~hosts could not be verified due to some kind of misconfiguration, \eg use of invalid certificates or certificates with missing issuer information.
Recall that we use OpenSSL trusted root certificates for validation.
Compared with other Web PKI studies we see in our focused sample of AA organizations relatively less invalid certificates compared to global average of $65\%$ as observed by Chung et al.~\cite{Chung2016} over the IPv4 space in 2016, or $\approx 13\%$ as measured by Durumeric et al.~\cite{Durumeric2013} for Alexa 1M top domain list in 2013.

\begin{table}
	\scriptsize
	\caption{DNS and Web PKI alongside assurance profiles}
	\begin{minipage}{\columnwidth}
		\setlength{\extrarowheight}{.25em}
		\begin{tabularx}{\columnwidth}{c c l c c X<{\centering} r}
			\toprule
			\multicolumn{2}{c}{DNS} & & \multicolumn{2}{c}{Certificate} & & \\
			\cmidrule{1-2} \cmidrule{4-5}
			\shortstack{Restricted\\delegation} & \shortstack{Supports\\DNSSEC} & & DV & O/EV & Assurance profile\textsuperscript{1} & \# Names\\
			\hline
			\cmark & \cmark & & -- & \cmark & \full & 29 ($\approx2\%$)\\
			\hline
			\cmark & \cmark & & \cmark & \xmark & \half & 11\\\dhline
			\xmark & \cmark & & -- & \cmark & \half & 2\\\dhline
			\cmark & \xmark & & -- & \cmark & \half & 132 \\\dhline
			\xmark & \xmark & & -- & \cmark & \half & 117 \\
			\dcline{1-6}\cline{7-7}
			\multicolumn{6}{r}{Total:} & 262 ($\approx20\%$)\\
			\hline
			\cmark & \xmark & & \cmark & \xmark & \none & 354\\\dhline
			\xmark & \xmark & & \cmark & \xmark & \none & 482\\\dhline
			\xmark & \cmark & & \cmark & \xmark & \none & 3\\\dhline
			\cmark & \cmark & & \xmark & \xmark & \none & 2\\\dhline
			\cmark & \xmark & & \xmark & \xmark & \none & 67\\\dhline
			\xmark & \cmark & & \xmark & \xmark & \none & 2\\\dhline
			\xmark & \xmark & & \xmark & \xmark & \none & 126\\
			\dcline{1-6}\cline{7-7}
			\multicolumn{6}{r}{Total:} & 1036 ($\approx78\%$)\\
			\hline
			\multicolumn{6}{r}{Grand Total:} & 1327 \\
			\bottomrule
		\end{tabularx}
		\smallskip
		
		{\scriptsize\textsuperscript{1}~\full~strong, \half~weak, \none~inadequate (see \autoref{tab:trust-matrix})}
		\label{tab:dns-tls-id}
	\end{minipage}
\end{table}

\begin{table}
	\scriptsize
	\caption{Validation types and assurance profiles per sector}
	\begin{minipage}{\columnwidth}
		\setlength{\extrarowheight}{.25em}
		\begin{tabularx}{\columnwidth}{X c c c c l >{\centering}p{15pt} >{\centering}p{15pt} p{15pt}<{\centering}}
			\toprule
			& \multicolumn{4}{c}{Certificate} & & \multicolumn{3}{c}{Assurance profile\textsuperscript{1}}\\
			\cline{2-5} \cline{7-9}
			Type & N/A & DV & OV & EV & & \full & \half & \none \\
			\hline
			Public Safety & 102 & 415 & 119 & 8 & & 10 & 120 & 514 \\\dhline
			Governmental & 73 & 318 & 102 & 6 & & 7 & 104 & 388 \\\dhline
			Law Enforcement & 21 & 110 & 31 & 0 & & 5 & 28 & 129 \\\dhline
			Military & 1 & 4 & 5 & 1 & & 6 & 3 & 2 \\\dhline
			Educational & 0 & 0 & 4 & 0 & & 0 & 4 & 0 \\\dhline
			Other & 0 & 3 & 3 & 1 & & 1 & 3 & 3\\
			\hline
			Total & 197 & 850 & 264 & 16 & & 29 & 262 & 1036 \\
			\bottomrule
		\end{tabularx}
		{\scriptsize\textsuperscript{1}~\full~strong, \half~weak, \none~inadequate (see \autoref{tab:trust-matrix})}
		\label{tab:tls-val-id}
	\end{minipage}
\end{table}

\autoref{tab:dns-tls-id} combines our findings from this Section and Section~\ref{sec:dns-analysis} to reveal different combinations of DNS and X.509 certificate characteristics, linked to different levels of assurance according to \autoref{tab:trust-matrix}. 
In \autoref{tab:tls-val-id}, we group our results by organization types.
Due to low penetration of DNSSEC, popularity of open TLDs, and pervasiveness of DV certificates among AAs (\S~\ref{sec:dns-analysis}), only about 22\% of AA are considered to be equipped against common threats to trustworthy communication.

\subsection{Historic X.509 Certificate Landscape}
The historic analysis of X.509 certificates collected from Certificate Transparency logs (see Section~\ref{sec:methodology}) helps us to gain a better understanding of security policy changes related to Alerting Authorities and CAs.
We span ten years.
It should be noted that the total number of organizations with publicly logged certificates changes for each year.
We consider this in the following and normalize the results either with respect to the number of organizations or total number of certificates valid per year.

\subsubsection{Certificate Authorities}
In addition to common regulations, certificate authorities implement and follow their own set of policies.
From the perspective of relying parties, \ie web users, such policies are opaque and as long as a CA is included in a user's trust store, it is considered trustworthy.
For the subscribers, however, these policies among other factors such as fees, offered certificate types, and operation costs are decisive in choosing an appropriate CA.

\begin{figure}
	\scriptsize
	\centering
	\includegraphics{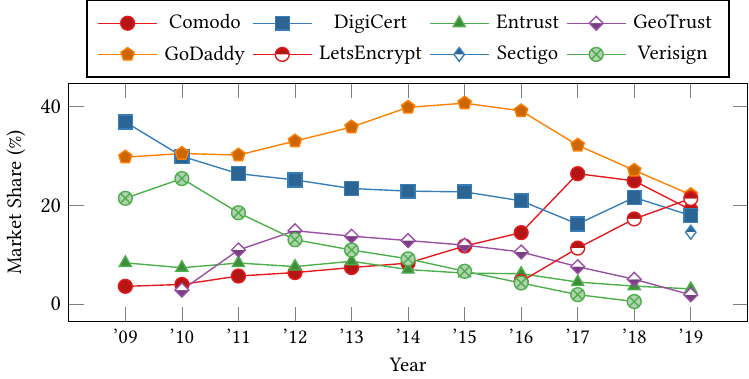}
	\caption{Market share of top CAs in the past decade}
	\label{fig:plot-ca-per-year-table}
\end{figure}

We focus our analysis on the list of top CAs with an average coverage of yearly 20~unique AA subscribers (hosts) in the last decade.
We use the term \textit{cover} to differentiate from issuance: if a host, for example, is issued a certificate by a CA valid from 2010 to 2013, we consider this host to be covered by that CA for 2010, 2011, 2012, and 2013.
Respectively, if a CA issues multiple short-lived certificates (\eg 90 days) for a host within a given year, we only count that host as covered once in that year by the issuing CA.
This would avoid the data skew in favor of issuers with lower certificate validity windows and higher certification rate per year.
It also should be noted that a single host can have certificates issued from different CAs.
\autoref{fig:plot-ca-per-year-table} depicts these findings in terms of relative market share development in the past decade (see \autoref{tab:total-certs-per-ca} in \hyperref[sec:app]{Appendix} for details).
Compared with the CA market share for the Alexa 1M top domain list throughout the last last decade~\cite{Holz2011,Durumeric2013,letsencrypt} we observe parallels, such as decline of GoDaddy's market share and rapid gains of Let's Encrypt, as well as discrepancies that cannot directly be explained due to dynamic nature of and fluctuations in the Alexa top list.

\autoref{fig:plot-ca-per-year-table} also highlights two factors evidently decisive for AAs in their choice of CA: convenience and cost factors.
GoDaddy, for example, which has been the market leader among AAs for about two thirds of the past decade, provides web hosting and domain name registration beside certification services in convenient packages; and Let's Encrypt, which has surged to the top in the short period after its public offering, offers automated DV certification at no cost.

\subsubsection{Validation types and assurance profiles}
\begin{figure}
	\scriptsize
	\begin{subfigure}{.48\columnwidth}
		\centering
		\includegraphics{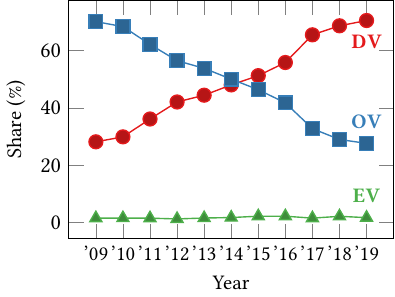}
		\caption{Certificate types}
	\end{subfigure}
	\begin{subfigure}{.48\columnwidth}
		\centering 
		\includegraphics{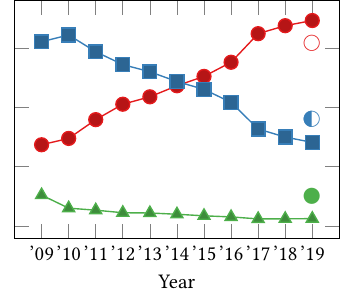}
		\caption{Assurance profiles}
	\end{subfigure}
	\caption{Development of certificate types and assurance profiles~(\S\ref{sec:trust}) in the past decade}
	\label{fig:plot-historic-cert-id-plot}
\end{figure}

In \autoref{subsec:webpkit-deployment}, we showed that currently only about 22\% of AAs honor security profiles that are resilient against threats to trustworthy communication~(see Tables~\ref{tab:dns-tls-id} and \ref{tab:tls-val-id}).
Historically, however, as depicted in \autoref{fig:plot-historic-cert-id-plot}, a higher share of alerting authorities provisioned for such measures.
When compared with the share of various certificate validation types (DV, OV, and EV), it becomes evident how the decreasing usage of OV certificates is directly proportional to the reduction of preferred assurance profiles.
At the same time the surging popularity of DV certificates has led to an increase in cases of what we consider as inadequately trustworthy~(no identification).
It should be noted that as our partial historical DNSSEC penetration statistics (collected through SecSpider\footnote{\url{https://secspider.net}}~\cite{osterweil2014shape}), covering $\approx 25\%$ of studied hosts, exhibits negligible fluctuation in DNSSEC penetration, we made a simple assumption that historic support for DNSSEC among AAs equals to its current penetration state (see \autoref{subsec:dnssec}).

\subsubsection{Certificate Sharing}
Except EV certificates, both DV and OV certificates allow wildcard names as subject alternative names (SAN) to avoid enumerating all FQDNs under the control of the certificate holder.
In practice, the SAN extension also allows sharing a certificate among different hosts.
For example, In 2019 the federal government was issued OV certificates with more than 600 SAN entries each.
Certificate sharing expands the attack surface and increases operational costs since if one of the hosts is compromised or the certificate is revoked, every other host also need to configured with a new certificate
(sometimes called ``fate-sharing'').

\begin{figure}
	\centering
	\includegraphics{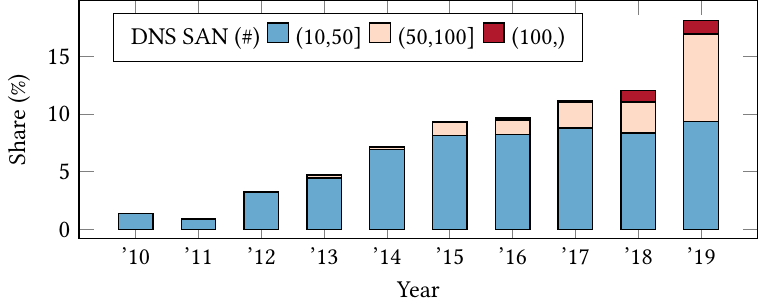}
	\caption{Share of host names represented by certificates with more than 10 unique SAN entries}
	\label{fig:san-per-aa-count}
\end{figure}

Multitenancy web hosting and security service providers (both public or government exclusive) are making use of shared certificates as depicted in \autoref{fig:san-per-aa-count}. It is worth noting that Let's Encrypt certificates only allow up to 100 DNS type SANs.
In our analysis, we also noticed an increasing number of certificate sharing among hosts which \emph{do not} belong to the same logical entity.
Most critically also among OV certificates where a service provider obtains a certificate under its name and lists the host name of its customers as SAN, practically defeating the identification purpose of the certificate.
At the time of writing, for example, we observe cases of such certificates listing SANs that obviously belong to separate entities, \eg \texttt{mo.gov}, \texttt{asap.farm}, and \texttt{incapsula.com} under the same certificate.
In this very specific case, records from the \href{https://web.archive.org/}{\emph{Wayback Machine}} archives show that \texttt{asap.farm} has previously belonged to Missouri Department of Agriculture~\cite{wayback-asapfarm} but it was never removed from the certificate as 
the domain name registration
was transferred to another entity.

\subsubsection{Certificate Validity}
A certificate is presumed valid if, among others, it is deployed within its validity period, is issued by a trustworthy CA, carries a valid signature, is bound to the correct subject name, and is not revoked~(see RFC 5280~\cite{RFC-5280}).
Checking revocation status often requires network transactions, and is the most expensive operation among aforementioned factors.  Thus in many cases it is either performed inadequately or ignored altogether by browsers (partly in favor of proprietary solutions)~\cite{Liu2015}.
Consequently, in the past years both CAs and browser vendors have been negotiating to cap and reduce certificate lifetimes~\cite{cab-ballot185,cab-ballot193,cab-ballotsc22} as an effort to reduce security risks due to misissued or revoked certificates.

\begin{figure}
	\scriptsize
	\includegraphics{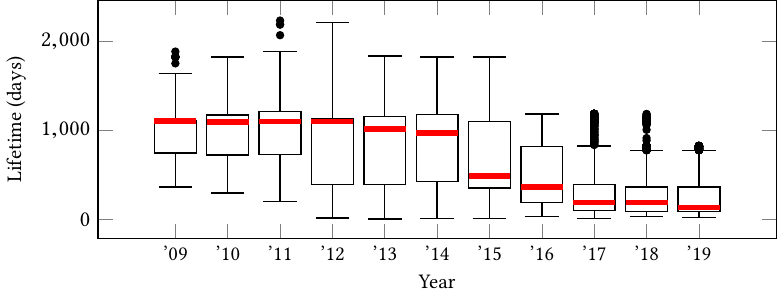}
	\caption{Validity distribution of logged certificates per year}
	\label{fig:plot-validity-per-year-boxplot}
\end{figure}

As depicted in \autoref{fig:plot-validity-per-year-boxplot}, the lifetime of certificates utilized by AAs has been constantly decreasing.
This trend can partly be attributed to consensus among CAs and browser vendors to reduce certificate lifetimes, but also due to rising popularity of CAs which are specialized on free and automated DV certificates such as Let's Encrypt (fixed lifetime of 90 days).
The median validity periods that we observe here are comparable with related works~\cite{Durumeric2013,Chung2016}, yet there are no recent studies that can corroborate the sharp decrease in validity periods from 2015 on.

\section{Related Work}
\label{sec:related}
To the best of our knowledge, this is the first study investigating how Alerting Authorities in the US (as part of broader critical infrastructure) implement measures to cater for trustworthy Web-based communication and service provision.
Previous research on trust in online emergency service provision mainly focuses on form and content and its relation to the perception of trustworthiness~\cite{Endsley2014,Busa2015,Hughes2015}, conception of trustworthy emergency communication and collaboration systems~\cite{Buscher2009,Palen2010} or simply best practices in building trust~\cite{Longstaff2008,PAHO2009}.
Although previous research has already highlighted how knowing who is behind an online emergency service impacts the trustworthiness of their respective services~\cite{Paton2007,Longstaff2008,Endsley2014}, we observe a research gap when it comes to evaluating the measures at one's disposal to reach this goal.
More specifically, the interplay of characteristics of domain names and X.509 certificates, \ie Assurance Profile (Section~\ref{sec:trust}).
has not been investigated to our best knowledge.
Respectively, we limit ourselves to an overview of related work which studies these technologies on their own.

\paragraph{Domain Namespace and DNSSEC}
The influence of a domain name on authenticating or at least recognizing the real-world entity behind that name has been investigated in terms of general trustworthiness associated with TLDs and impersonation of trusted entities through domain name masquerading.
Walther, Wang, and Loh~\cite{Walther2004} examine how choice of TLD can positively impact the credibility of health websites.
Seckler~\etal~\cite{Seckler2015} investigate how a relevant domain name, \eg a known TLD, can positively enforce familiarity and in turn increase trust.
Similarly, a yearly report~\cite{globaltrend-18} commissioned by the Public Internet Registry 
examines the trustworthiness of select TLDs among NGO donors.

A closely related topic is how the domain namespace of malicious websites is structured and operated.
Korczynski~\etal~\cite{Korczynski2018} show how low pricing and registration barriers alongside the possibility of bulk registration is an enabler for malicious actors to migrate to new gTLDs.
In a longitudinal study of typosquatting, Agten~\etal~\cite{Agten2015} reveals how registration fees and registry policies can attract or deter malicious actors; practically determining the credibility of such TLDs (the top three most abused TLDs in the world are new gTLDs~\cite{spamhaus20}).
And Antonakakis~\etal~\cite{Antonakakis2010} introduce a reputation system for DNS to detect malicious domain names.
Different studies show how scammers try to impersonate other entities by partly or fully integrating legitimate domain names in their own domain names~\cite{Agten2015,Kintis2017,Tian2018,Roberts2019} or even by using homonymous names using internationalized domain names~\cite{Suzuki2019}.

With regard to namespace security, studies in the past pinpoint a relatively low DNSSEC penetration due to various factors ranging from lack of support by local resolvers to server misconfigurations~\cite{Osterweil2008,HaoYang2011,Lian2013,Chung2017} despite more than 90\% of all TLDs being signed and supporting DNSSEC~\cite{dnssec-tld}.
The prevalence of DNSSEC among various types of organizations, such as educational, military, commercial, \etc has not been subject of study to determine if there is a correlation between field of operation and sensibility for DNS security measures.
The only exception is the fine-grained, \ie including second level domains, regular analysis of DNSSEC deployment among select governmental agencies within the \texttt{.gov} namespace, educational institutions, and industry in the US~\cite{dnssec-gov,Rose2012}.

\paragraph{Web PKI}
Throughout the years, various measurements have characterized X.509 certificates in use over the Internet in terms of validity, issuing CAs, key strength, \etc~\cite{Mishari2009,Holz2011,Durumeric2013,Chung2016}.
Among these, Mishari~\etal~\cite{Mishari2009} investigate the difference between certificates of legitimate and fraudulent websites.
The study by Holz~\etal~\cite{Holz2011} has the advantage of being performed from different vantage points spread over the world.
The measurements by Durumeric~\etal~\cite{Durumeric2013} is noteworthy as it goes beyond mere X.509 certificate analysis and investigates the dependencies among root and intermediate CAs, their market share, and the characteristics of respective certificates.
And finally, the measurements performed by Chung~\etal~\cite{Chung2016} aim to understand why a majority of certificates advertised over IPv4 are invalid.
It should be noted that except the last study, the others have been carried out before major changes in the Web~PKI occurred, such as various mergers and the public launch of Let's~Encrypt~\cite{letsencrypt}.
Furthermore, the findings from these studies exhibit different characteristics of sample sets, which are either too limited (\eg Alexa~Top~1M) or too broad (\eg IPv4 space).
Those differences do not allow for statistical inference and comparison with our observations.

In a recent study, which is most closely related to our work, Singanamalla~\etal~\cite{10.1145/3419394.3423645} measure the adoption of https at government websites.
Using primarily an automated, keyword-based matching to collect domain names this approach is prone to false positives and makes comparison to our work infeasible.
Also, this work does not relate to Assurance Profiles, which we introduce in our study.

Specifically related to the topic of our work are studies which investigate the trustworthiness of CAs in general and their policies specially in enabling fraud and impersonation.
Delignat-Lavaud~\textit{et al.} \cite{Delignat-Lavaud2014}, for example, investigate the conformance of CAs to the CA/Browser Forum guidelines, which in turn can influence trustworthiness of a CA.
Others have defined various metrics to qualify~\cite{Chadwick2001,Fadai2015} or quantify trustworthiness of CAs~\cite{Heinl2019} beyond technical measures.
In a recent study Schwittmann, Wander and Weis~\cite{Schwittmann2019}, similar to Brandt~\etal~\cite{Brandt2018}, exhibit how various CAs are susceptible to attacks on DV certification processes that can practically lead to domain impersonation.
Roberts~\etal~\cite{Roberts2019} studies which CAs are responsible for issuing DV certificates to malicious target-embedded domains.

\section{Key Findings and Discussions}\label{sec:discus}

Our results draw a rather alarming picture of the current online emergency management landscape regarding trustworthy communication.
Along the line of our key findings, we discuss the possible reasons for the observed deficiencies and suggest alternatives.

\paragraph{Only about 22\% of AAs deploy sufficient identification}
Identification, as discussed in \autoref{sec:trust}, succeeds over multiple factors, which are only insufficiently attended to by AAs:
only about half of AAs have their dedicated names and as such cannot obtain exclusive X.509 certificates as proof of identity (as these are bound to domain names) while a majority of $\approx 78\%$ fail to provide any valid certificate or just DV certificates which lack identification information.
The majority of organizations opt for generic TLDs which simplifies name spoofing and phishing as precursors of impersonation.
Additionally, the minuscule penetration rate of DNSSEC provides another attack surface by poisoning DNS records and misdirecting users to malicious websites.
Alerting Authorities should at best be located under restricted namespaces as an additional factor of recognizably and assurance, have at least their own subdomains instead of being subsumed in the path segment of a URL, secure their namespace using DNSSEC, and provide OV/EV certificates as definitive proof of identity.

\paragraph{Less than 4\% of AAs offer secure name resolution}
Securing domain names is seemingly a non-priority for investigated organizations as the low penetration rate of DNSSEC suggests.
Insecure DNS not only can cause misdirection from authentic websites, but also DV certificate misissuance~\cite{Brandt2018,Schwittmann2019} which impacts both identification and session security.
Although DNSSEC suffers low deployment on the global scale in general, it is an indispensable component in securing emergency communication as part of the broader critical infrastructure.
Yet, it should be noted that in some cases due to lack of support registrants are forced to abandon DNSSEC in favor of other factors, \eg registering under a \texttt{.us} locality name for which there is, surprisingly, no DNSSEC support (see \autoref{fig:dnssec-per-state-map}).
We also note that although domain names under \texttt{.gov} namespaces are mandated to use DNSSEC~\cite{Technology2009}, the low support for DNSSEC has its roots in operational and organizational mismanagement rather than technical issues.

\paragraph{DV certificates dominate transaction security}
The popularity of domain validation certificates combined with low penetration of DNSSEC represents an attack surface that can compromise session security through certificate missisuance and monkey-in-the-middle attacks.
If DV is indispensable for some, we encourage the stakeholders to reconsider semantically equivalent alternative of TLSA domain issued certificates (DANE EE) as they provide higher resilience against spoofing in contrast to DV certificates~\cite{Schwittmann2019}.
In general, DANE can be used to remove ambiguity regarding public keys and responsible CAs for a domain name~\cite{osterweil2014shape}.

\paragraph{Fate-sharing is on the rise}
The lack of dedicated domain names and an increase of certificate sharing in multitenancy settings represents worrisome and de facto unnecessary dependencies, which both can expand the attack surface~\cite{osterweil2014shape} and can cause instabilities in the future.
Regarding shared certificates, we suggest abandoning them completely and also encourage CAs to avoid issuing OV certificates for service providers without ensuring that all the listed subject alternative names belong to the same organization.

\paragraph{Convenience and cost impact security preferences}
We observe 15\% of AAs providing none or invalid certificates.
This can be traced back to carelessness regarding the Web PKI trust model (self-signed certificates) or additional (not only financial) configuration~\cite{Krombholz2017} and certification costs.
Rapid growth of Let's Encrypt with its fully automated certificate issuance and renewal is an indication of how the aforementioned factors influence the decision for choosing an appropriate CA.
Similarly, we measure less than 45\% represented under restricted namespaces.
In contrast to gTLDs, higher registration fees or bureaucratic hurdles, and longer delegation processing times are among discouraging factors, which call for governmental support and can effectively be addressed by policy-makers through price caps and easier access for eligible organizations which fulfill the strict requirements.

\paragraph{Responsibilities beyond Alerting Authorities}
The scope of trustworthy communication goes beyond our investigations and extends to consumers as well as infrastructure operators such as CAs, ISPs, and browser vendors.
There is still a gap between CA practices and guidelines~\cite{Delignat-Lavaud2014}, some automated DV certification services are susceptible to impersonation attacks~\cite{Schwittmann2019,Roberts2019}, and some root CAs do not restrict certification scope for their intermediate CAs~\cite{Durumeric2013}.
DNS registrars not always offer DNSSEC by default or free of cost~\cite{Chung-dnssec-2017} 
and only a minority of ISPs bother to operate DNSSEC-aware recursive resolvers that properly 
verify signed DNS records~\cite{Wander2013,Chung2017}.
Browser vendors should also provide better security usability by avoiding confusing SSL/TLS warnings~\cite{Akhawe2013}, improve instead of abandoning visual cues for different certificate types~\cite{Felt2019,chrome-ev-19,firefox-ev-19}, and start offering alternative CA trustworthiness assessment measures beyond the standard binary trust model~\cite{Chadwick2001,Fadai2015,Heinl2019}.
Finally, users should be educated in better understanding the semantics of domain names~\cite{roberts20} and web PKI certificates and their practical use and ramifications.

\section{Conclusion and Outlook}
In this paper, we conceptualized a threat model for trustworthy communication in emergency management and analyzed the lack of common technologies, DNS(SEC) and Web PKI, to mitigate threats to identification, resolution, and content manipulation or eavesdropping.
We provided an overview of how Alerting Authorities~(AA) in the US are structured within the domain namespace, how widespread is DNSSEC in securing their domain names, and how Web PKI is used for authentication and data security.
We uncovered deficiencies and discussed alternatives while emphasizing that respective solutions are not necessarily technical but operational as well as political.
Protecting critical infrastructure for emergency communication and public safety entails addressing operational and policy challenges on national and international levels and calls for commitment of all stakeholders from service providers to intermediate infrastructure operators and browser vendors alongside policy-makers.

In the future, this work can be extended beyond the US territory while providing a comparison basis for other countries.
Furthermore, other technologies can be accommodated in our assurance profiles.
Finally, the role of intermediate infrastructure and further dependency structures can be investigated in~depth.

\paragraph{Data Disclosure}
We provide a browser that presents the assurance profile of each Alerting Authority and additional accompanying material on \href{https://aa.secnow.net/}{\texttt{https://aa.secnow.net}}.
Our toolchain and collected data are published under \href{https://doi.org/10.5281/zenodo.4300946}{\texttt{doi:10.5281/zenodo.4300946}}.

\paragraph{Ethical Considerations}
We informed Alerting Authorities about their assurance profiles to raise awareness for improvements.

\paragraph{Acknowledgments}
This work was supported in parts by the German Federal Ministry of Education and Research (BMBF) within the project \emph{Deutsches Internet-Institut} (grant no. \textit{16DII111}).

\bibliographystyle{ACM-Reference-Format}
\bibliography{sigproc,rfcs}


\begin{thebibliography}{91}


\ifx \showCODEN    \undefined \def \showCODEN     #1{\unskip}     \fi
\ifx \showDOI      \undefined \def \showDOI       #1{#1}\fi
\ifx \showISBNx    \undefined \def \showISBNx     #1{\unskip}     \fi
\ifx \showISBNxiii \undefined \def \showISBNxiii  #1{\unskip}     \fi
\ifx \showISSN     \undefined \def \showISSN      #1{\unskip}     \fi
\ifx \showLCCN     \undefined \def \showLCCN      #1{\unskip}     \fi
\ifx \shownote     \undefined \def \shownote      #1{#1}          \fi
\ifx \showarticletitle \undefined \def \showarticletitle #1{#1}   \fi
\ifx \showURL      \undefined \def \showURL       {\relax}        \fi
\providecommand\bibfield[2]{#2}
\providecommand\bibinfo[2]{#2}
\providecommand\natexlab[1]{#1}
\providecommand\showeprint[2][]{arXiv:#2}

\bibitem[\protect\citeauthoryear{3rd}{3rd}{2011}]%
        {RFC-6066}
\bibfield{author}{\bibinfo{person}{D.~Eastlake 3rd}.}
  \bibinfo{year}{2011}\natexlab{}.
\newblock \bibinfo{booktitle}{\emph{{Transport Layer Security (TLS) Extensions:
  Extension Definitions}}}.
\newblock \bibinfo{type}{RFC} 6066. \bibinfo{institution}{IETF}.
\newblock


\bibitem[\protect\citeauthoryear{Aas, Rescorla, Schoen, Warren, Barnes, Case,
  Durumeric, Eckersley, Flores-L{\'{o}}pez, Halderman, Hoffman-Andrews, and
  Kasten}{Aas et~al\mbox{.}}{2019}]%
        {letsencrypt}
\bibfield{author}{\bibinfo{person}{Josh Aas}, \bibinfo{person}{Eric Rescorla},
  \bibinfo{person}{Seth Schoen}, \bibinfo{person}{Brad Warren},
  \bibinfo{person}{Richard Barnes}, \bibinfo{person}{Benton Case},
  \bibinfo{person}{Zakir Durumeric}, \bibinfo{person}{Peter Eckersley},
  \bibinfo{person}{Alan Flores-L{\'{o}}pez}, \bibinfo{person}{J~Alex
  Halderman}, \bibinfo{person}{Jacob Hoffman-Andrews}, {and}
  \bibinfo{person}{James Kasten}.} \bibinfo{year}{2019}\natexlab{}.
\newblock \showarticletitle{{Let's Encrypt: An Automated Certificate Authority
  to Encrypt the Entire Web}}. In \bibinfo{booktitle}{\emph{Proc. of the 2019
  ACM SIGSAC CCS}}. \bibinfo{publisher}{ACM Press}, \bibinfo{address}{New York,
  NY, USA}, \bibinfo{pages}{2473--2487}.
\newblock
\showISBNx{9781450367479}


\bibitem[\protect\citeauthoryear{Administration}{Administration}{2020}]%
        {usanlytics}
\bibfield{author}{\bibinfo{person}{U.S. General~Services Administration}.}
  \bibinfo{year}{2020}\natexlab{}.
\newblock \bibinfo{title}{Digital Analytics Program.}
\newblock
\newblock
\urldef\tempurl%
\url{https://analytics.usa.gov/}
\showURL{%
\tempurl}


\bibitem[\protect\citeauthoryear{Agten, Joosen, Piessens, and
  Nikiforakis}{Agten et~al\mbox{.}}{2015}]%
        {Agten2015}
\bibfield{author}{\bibinfo{person}{Pieter Agten}, \bibinfo{person}{Wouter
  Joosen}, \bibinfo{person}{Frank Piessens}, {and} \bibinfo{person}{Nick
  Nikiforakis}.} \bibinfo{year}{2015}\natexlab{}.
\newblock \showarticletitle{{Seven Months' Worth of Mistakes: A Longitudinal
  Study of Typosquatting Abuse}}. In \bibinfo{booktitle}{\emph{Proc. of the
  2015 NDSS}}. \bibinfo{publisher}{Internet Society}, \bibinfo{address}{Reston,
  VA, USA}, \bibinfo{pages}{8--11}.
\newblock
\showISBNx{1-891562-38-X}


\bibitem[\protect\citeauthoryear{Akhawe and Felt}{Akhawe and Felt}{2013}]%
        {Akhawe2013}
\bibfield{author}{\bibinfo{person}{Devdatta Akhawe} {and}
  \bibinfo{person}{Adrienne~Porter Felt}.} \bibinfo{year}{2013}\natexlab{}.
\newblock \showarticletitle{{Alice in warningland: A large-scale field study of
  browser security warning effectiveness}}. In \bibinfo{booktitle}{\emph{Proc.
  of 22nd {USENIX} Security Symposium}}. \bibinfo{publisher}{{USENIX}
  Association}, \bibinfo{pages}{257--272}.
\newblock
\showISBNx{978-1-931971-03-4}


\bibitem[\protect\citeauthoryear{Antonakakis, Perdisci, Dagon, Lee, and
  Feamster}{Antonakakis et~al\mbox{.}}{2010}]%
        {Antonakakis2010}
\bibfield{author}{\bibinfo{person}{Manos Antonakakis}, \bibinfo{person}{Roberto
  Perdisci}, \bibinfo{person}{David Dagon}, \bibinfo{person}{Wenke Lee}, {and}
  \bibinfo{person}{Nick Feamster}.} \bibinfo{year}{2010}\natexlab{}.
\newblock \showarticletitle{{Building a dynamic reputation system for DNS}}. In
  \bibinfo{booktitle}{\emph{Proc. of the 19th USENIX Security Symposium}}.
  \bibinfo{publisher}{{USENIX} Association}, \bibinfo{pages}{273--289}.
\newblock
\showISBNx{9781931971775}


\bibitem[\protect\citeauthoryear{Arends, Austein, Larson, Massey, and
  Rose}{Arends et~al\mbox{.}}{2005a}]%
        {RFC-4033}
\bibfield{author}{\bibinfo{person}{R. Arends}, \bibinfo{person}{R. Austein},
  \bibinfo{person}{M. Larson}, \bibinfo{person}{D. Massey}, {and}
  \bibinfo{person}{S. Rose}.} \bibinfo{year}{2005}\natexlab{a}.
\newblock \bibinfo{booktitle}{\emph{{DNS Security Introduction and
  Requirements}}}.
\newblock \bibinfo{type}{RFC} 4033. \bibinfo{institution}{IETF}.
\newblock


\bibitem[\protect\citeauthoryear{Arends, Austein, Larson, Massey, and
  Rose}{Arends et~al\mbox{.}}{2005b}]%
        {RFC-4035}
\bibfield{author}{\bibinfo{person}{R. Arends}, \bibinfo{person}{R. Austein},
  \bibinfo{person}{M. Larson}, \bibinfo{person}{D. Massey}, {and}
  \bibinfo{person}{S. Rose}.} \bibinfo{year}{2005}\natexlab{b}.
\newblock \bibinfo{booktitle}{\emph{{Protocol Modifications for the DNS
  Security Extensions}}}.
\newblock \bibinfo{type}{RFC} 4035. \bibinfo{institution}{IETF}.
\newblock


\bibitem[\protect\citeauthoryear{Arends, Austein, Larson, Massey, and
  Rose}{Arends et~al\mbox{.}}{2005c}]%
        {RFC-4034}
\bibfield{author}{\bibinfo{person}{R. Arends}, \bibinfo{person}{R. Austein},
  \bibinfo{person}{M. Larson}, \bibinfo{person}{D. Massey}, {and}
  \bibinfo{person}{S. Rose}.} \bibinfo{year}{2005}\natexlab{c}.
\newblock \bibinfo{booktitle}{\emph{{Resource Records for the DNS Security
  Extensions}}}.
\newblock \bibinfo{type}{RFC} 4034. \bibinfo{institution}{IETF}.
\newblock


\bibitem[\protect\citeauthoryear{{Arroyo Barrantes}, Rodriguez, and
  P{\'{e}}rez}{{Arroyo Barrantes} et~al\mbox{.}}{2009}]%
        {PAHO2009}
\bibfield{editor}{\bibinfo{person}{Susana {Arroyo Barrantes}},
  \bibinfo{person}{Martha Rodriguez}, {and} \bibinfo{person}{Ricardo
  P{\'{e}}rez}} (Eds.). \bibinfo{year}{2009}\natexlab{}.
\newblock \bibinfo{booktitle}{\emph{{Information Management and Communication
  in Emergencies and Disasters}}}.
\newblock \bibinfo{publisher}{Pan American Health Organization},
  \bibinfo{address}{Washington D.C., USA}.
\newblock


\bibitem[\protect\citeauthoryear{Association and Wixted}{Association and
  Wixted}{2018}]%
        {national2018nfpa}
\bibfield{author}{\bibinfo{person}{National Fire~Protection Association} {and}
  \bibinfo{person}{M.T. Wixted}.} \bibinfo{year}{2018}\natexlab{}.
\newblock \bibinfo{booktitle}{\emph{NFPA 1600, Standard on Continuity,
  Emergency, and Crisis Management, 2019 Edition}}.
\newblock \bibinfo{publisher}{National Fire Protection Association},
  \bibinfo{address}{Quincy, MA, USA}.
\newblock
\showISBNx{9781455922093}


\bibitem[\protect\citeauthoryear{Atkins and Austein}{Atkins and
  Austein}{2004}]%
        {RFC-3833}
\bibfield{author}{\bibinfo{person}{D. Atkins} {and} \bibinfo{person}{R.
  Austein}.} \bibinfo{year}{2004}\natexlab{}.
\newblock \bibinfo{booktitle}{\emph{{Threat Analysis of the Domain Name System
  (DNS)}}}.
\newblock \bibinfo{type}{RFC} 3833. \bibinfo{institution}{IETF}.
\newblock


\bibitem[\protect\citeauthoryear{Blanchard}{Blanchard}{2008}]%
        {Blanchard2008}
\bibfield{author}{\bibinfo{person}{B.~Wayne Blanchard}.}
  \bibinfo{year}{2008}\natexlab{}.
\newblock \bibinfo{title}{{Guide To Emergency Management and Related Terms,
  Definitions, Concepts, Acronyms, Organizations, Programs, Guidance, Executive
  Orders {\&} Legislation}}.
\newblock , \bibinfo{numpages}{1366}~pages.
\newblock


\bibitem[\protect\citeauthoryear{Brandt, Dai, Klein, Shulman, and
  Waidner}{Brandt et~al\mbox{.}}{2018}]%
        {Brandt2018}
\bibfield{author}{\bibinfo{person}{Markus Brandt}, \bibinfo{person}{Tianxiang
  Dai}, \bibinfo{person}{Amit Klein}, \bibinfo{person}{Haya Shulman}, {and}
  \bibinfo{person}{Michael Waidner}.} \bibinfo{year}{2018}\natexlab{}.
\newblock \showarticletitle{{Domain Validation++ For MitM-Resilient PKI}}. In
  \bibinfo{booktitle}{\emph{Proc. of the 2018 ACM SIGSAC}}.
  \bibinfo{publisher}{ACM Press}, \bibinfo{address}{New York, NY, USA},
  \bibinfo{pages}{2060--2076}.
\newblock
\showISBNx{9781450356930}
\showISSN{15437221}


\bibitem[\protect\citeauthoryear{Brennen, Simon, Howard, and Nielsen}{Brennen
  et~al\mbox{.}}{2020}]%
        {ox-misinfo20}
\bibfield{author}{\bibinfo{person}{J.~Scott Brennen}, \bibinfo{person}{Felix
  Simon}, \bibinfo{person}{Philip~N. Howard}, {and}
  \bibinfo{person}{Rasmus~Kleis Nielsen}.} \bibinfo{year}{2020}\natexlab{}.
\newblock \bibinfo{title}{Types, sources, and claims of COVID-19
  misinformation}.
\newblock
\newblock
\urldef\tempurl%
\url{https://reutersinstitute.politics.ox.ac.uk/types-sources-and-claims-covid-19-misinformation}
\showURL{%
\tempurl}


\bibitem[\protect\citeauthoryear{Burke}{Burke}{2020}]%
        {nbc-crash20}
\bibfield{author}{\bibinfo{person}{Minyvonne Burke}.}
  \bibinfo{year}{2020}\natexlab{}.
\newblock \bibinfo{title}{Coronavirus: State unemployment websites crash as
  applications surge}.
\newblock
\newblock
\urldef\tempurl%
\url{https://www.nbcnews.com/news/us-news/coronavirus-state-unemployment-websites-crash-applications-surge-n1162731}
\showURL{%
\tempurl}


\bibitem[\protect\citeauthoryear{Bus{\`{a}}, Musacchio, Finan, and
  Stillwater}{Bus{\`{a}} et~al\mbox{.}}{2015}]%
        {Busa2015}
\bibfield{author}{\bibinfo{person}{Maria~Grazia Bus{\`{a}}},
  \bibinfo{person}{Maria~Teresa Musacchio}, \bibinfo{person}{Shane Finan},
  {and} \bibinfo{person}{Cilian~Fennel Stillwater}.}
  \bibinfo{year}{2015}\natexlab{}.
\newblock \showarticletitle{{Trust-building through social media communications
  in disaster management}}. In \bibinfo{booktitle}{\emph{Companion Proc. of the
  24th ACM WWW}}. \bibinfo{publisher}{ACM}, \bibinfo{address}{New York, NY,
  USA}, \bibinfo{pages}{1179--1184}.
\newblock
\showISBNx{9781450334730}


\bibitem[\protect\citeauthoryear{B{\"{u}}scher, {Holst Mogensen}, and
  Kristensen}{B{\"{u}}scher et~al\mbox{.}}{2009}]%
        {Buscher2009}
\bibfield{author}{\bibinfo{person}{Monika B{\"{u}}scher},
  \bibinfo{person}{Preben {Holst Mogensen}}, {and} \bibinfo{person}{Margit
  Kristensen}.} \bibinfo{year}{2009}\natexlab{}.
\newblock \showarticletitle{{When and How (Not) to Trust It? Supporting Virtual
  Emergency Teamwork}}.
\newblock \bibinfo{journal}{\emph{International Journal of Information Systems
  for Crisis Response and Management}} \bibinfo{volume}{1}, \bibinfo{number}{2}
  (\bibinfo{date}{apr} \bibinfo{year}{2009}), \bibinfo{pages}{1--15}.
\newblock
\showISSN{1937-9390}


\bibitem[\protect\citeauthoryear{Chadwick and Basden}{Chadwick and
  Basden}{2001}]%
        {Chadwick2001}
\bibfield{author}{\bibinfo{person}{David~W. Chadwick} {and}
  \bibinfo{person}{Andrew Basden}.} \bibinfo{year}{2001}\natexlab{}.
\newblock \showarticletitle{{Evaluating trust in a public key certification
  authority}}.
\newblock \bibinfo{journal}{\emph{Computers and Security}}
  \bibinfo{volume}{20}, \bibinfo{number}{7} (\bibinfo{year}{2001}),
  \bibinfo{pages}{592--611}.
\newblock
\showISSN{01674048}


\bibitem[\protect\citeauthoryear{Chauhan and Hughes}{Chauhan and
  Hughes}{2017}]%
        {Chauhan2017}
\bibfield{author}{\bibinfo{person}{Apoorva Chauhan} {and}
  \bibinfo{person}{Amanda~Lee Hughes}.} \bibinfo{year}{2017}\natexlab{}.
\newblock \showarticletitle{{Providing Online Crisis Information}}. In
  \bibinfo{booktitle}{\emph{Proc. of the 2017 CHI}}. \bibinfo{publisher}{ACM
  Press}, \bibinfo{address}{New York, NY, USA}, \bibinfo{pages}{3151--3162}.
\newblock
\showISBNx{9781450346559}


\bibitem[\protect\citeauthoryear{Chung, Liu, Choffnes, Levin, Maggs, Mislove,
  and Wilson}{Chung et~al\mbox{.}}{2016}]%
        {Chung2016}
\bibfield{author}{\bibinfo{person}{Taejoong Chung}, \bibinfo{person}{Yabing
  Liu}, \bibinfo{person}{David Choffnes}, \bibinfo{person}{Dave Levin},
  \bibinfo{person}{Bruce~MacDowell Maggs}, \bibinfo{person}{Alan Mislove},
  {and} \bibinfo{person}{Christo Wilson}.} \bibinfo{year}{2016}\natexlab{}.
\newblock \showarticletitle{{Measuring and Applying Invalid SSL Certificates:
  The Silent Majority}}. In \bibinfo{booktitle}{\emph{Proc. of the ACM IMC
  '16}}. \bibinfo{publisher}{ACM Press}, \bibinfo{address}{New York, NY, USA},
  \bibinfo{pages}{527--541}.
\newblock
\showISBNx{9781450345262}


\bibitem[\protect\citeauthoryear{Chung, {Van Rijswijk-Deij}, Chandrasekaran,
  Choffnes, Levin, Maggs, Mislove, and Wilson}{Chung et~al\mbox{.}}{2017a}]%
        {Chung2017}
\bibfield{author}{\bibinfo{person}{Taejoong Chung}, \bibinfo{person}{Roland
  {Van Rijswijk-Deij}}, \bibinfo{person}{Balakrishnan Chandrasekaran},
  \bibinfo{person}{David Choffnes}, \bibinfo{person}{Dave Levin},
  \bibinfo{person}{Bruce~M. Maggs}, \bibinfo{person}{Alan Mislove}, {and}
  \bibinfo{person}{Christo Wilson}.} \bibinfo{year}{2017}\natexlab{a}.
\newblock \showarticletitle{{A longitudinal, end-to-end view of the DNSSEC
  ecosystem}}. In \bibinfo{booktitle}{\emph{Proc. of the 26th USENIX Security
  Symposium}}. \bibinfo{publisher}{USENIX Association},
  \bibinfo{pages}{1307--1322}.
\newblock
\showISBNx{9781931971409}


\bibitem[\protect\citeauthoryear{Chung, van Rijswijk-Deij, Choffnes, Levin,
  Maggs, Mislove, and Wilson}{Chung et~al\mbox{.}}{2017b}]%
        {Chung-dnssec-2017}
\bibfield{author}{\bibinfo{person}{Taejoong Chung}, \bibinfo{person}{Roland van
  Rijswijk-Deij}, \bibinfo{person}{David Choffnes}, \bibinfo{person}{Dave
  Levin}, \bibinfo{person}{Bruce~M. Maggs}, \bibinfo{person}{Alan Mislove},
  {and} \bibinfo{person}{Christo Wilson}.} \bibinfo{year}{2017}\natexlab{b}.
\newblock \showarticletitle{Understanding the Role of Registrars in DNSSEC
  Deployment}. In \bibinfo{booktitle}{\emph{Proc. of the ACM IMC '17}}.
  \bibinfo{publisher}{ACM}, \bibinfo{address}{New York, NY, USA},
  \bibinfo{pages}{369–383}.
\newblock
\showISBNx{9781450351188}


\bibitem[\protect\citeauthoryear{Condon and Robinson}{Condon and
  Robinson}{2014}]%
        {Condon2014}
\bibfield{author}{\bibinfo{person}{Sherri~L. Condon} {and}
  \bibinfo{person}{Jason~R. Robinson}.} \bibinfo{year}{2014}\natexlab{}.
\newblock \showarticletitle{{Communication media use in emergency response
  management}}. In \bibinfo{booktitle}{\emph{ISCRAM 2014 Conference
  Proceedings}}. \bibinfo{publisher}{ISCRAM}, \bibinfo{pages}{687--696}.
\newblock
\showISBNx{9780692211946}


\bibitem[\protect\citeauthoryear{Cooper and Postel}{Cooper and Postel}{1993}]%
        {RFC-1480}
\bibfield{author}{\bibinfo{person}{A. Cooper} {and} \bibinfo{person}{J.
  Postel}.} \bibinfo{year}{1993}\natexlab{}.
\newblock \bibinfo{booktitle}{\emph{{The US Domain}}}.
\newblock \bibinfo{type}{RFC} 1480. \bibinfo{institution}{IETF}.
\newblock


\bibitem[\protect\citeauthoryear{Cooper, Santesson, Farrell, Boeyen, Housley,
  and Polk}{Cooper et~al\mbox{.}}{2008}]%
        {RFC-5280}
\bibfield{author}{\bibinfo{person}{D. Cooper}, \bibinfo{person}{S. Santesson},
  \bibinfo{person}{S. Farrell}, \bibinfo{person}{S. Boeyen},
  \bibinfo{person}{R. Housley}, {and} \bibinfo{person}{W. Polk}.}
  \bibinfo{year}{2008}\natexlab{}.
\newblock \bibinfo{booktitle}{\emph{{Internet X.509 Public Key Infrastructure
  Certificate and Certificate Revocation List (CRL) Profile}}}.
\newblock \bibinfo{type}{RFC} 5280. \bibinfo{institution}{IETF}.
\newblock


\bibitem[\protect\citeauthoryear{Delignat-Lavaud, Abadi, Birrell, Mironov,
  Wobber, and Xie}{Delignat-Lavaud et~al\mbox{.}}{2014}]%
        {Delignat-Lavaud2014}
\bibfield{author}{\bibinfo{person}{Antoine Delignat-Lavaud},
  \bibinfo{person}{Mart{\'{i}}n Abadi}, \bibinfo{person}{Andrew Birrell},
  \bibinfo{person}{Ilya Mironov}, \bibinfo{person}{Ted Wobber}, {and}
  \bibinfo{person}{Yinglian Xie}.} \bibinfo{year}{2014}\natexlab{}.
\newblock \showarticletitle{{Web PKI: Closing the Gap between Guidelines and
  Practices}}. In \bibinfo{booktitle}{\emph{Proc. of the 2014 NDSS}}.
  \bibinfo{publisher}{Internet Society}, \bibinfo{address}{Reston, VA, USA},
  \bibinfo{pages}{23--26}.
\newblock
\showISBNx{1-891562-35-5}


\bibitem[\protect\citeauthoryear{Durumeric, Kasten, Bailey, and
  Halderman}{Durumeric et~al\mbox{.}}{2013}]%
        {Durumeric2013}
\bibfield{author}{\bibinfo{person}{Zakir Durumeric}, \bibinfo{person}{James
  Kasten}, \bibinfo{person}{Michael Bailey}, {and} \bibinfo{person}{J.~Alex
  Halderman}.} \bibinfo{year}{2013}\natexlab{}.
\newblock \showarticletitle{{Analysis of the HTTPS certificate ecosystem}}. In
  \bibinfo{booktitle}{\emph{Proc. of the ACM IMC '13}}. \bibinfo{publisher}{ACM
  Press}, \bibinfo{pages}{291--304}.
\newblock
\showISBNx{9781450319539}


\bibitem[\protect\citeauthoryear{Endsley, Wu, and Reep}{Endsley
  et~al\mbox{.}}{2014}]%
        {Endsley2014}
\bibfield{author}{\bibinfo{person}{Tristan Endsley}, \bibinfo{person}{Yu Wu},
  {and} \bibinfo{person}{James Reep}.} \bibinfo{year}{2014}\natexlab{}.
\newblock \showarticletitle{{The source of the story: Evaluating the
  credibility of crisis information sources}}.
\newblock \bibinfo{journal}{\emph{ISCRAM 2014 Conference Proceedings}}
  \bibinfo{volume}{1}, \bibinfo{number}{1} (\bibinfo{year}{2014}),
  \bibinfo{pages}{160--164}.
\newblock
\showISBNx{9780692211946}


\bibitem[\protect\citeauthoryear{Fadai, Schrittwieser, Kieseberg, and
  Mulazzani}{Fadai et~al\mbox{.}}{2015}]%
        {Fadai2015}
\bibfield{author}{\bibinfo{person}{Tariq Fadai}, \bibinfo{person}{Sebastian
  Schrittwieser}, \bibinfo{person}{Peter Kieseberg}, {and}
  \bibinfo{person}{Martin Mulazzani}.} \bibinfo{year}{2015}\natexlab{}.
\newblock \showarticletitle{Trust Me, I’m a Root CA! Analyzing SSL Root CAs
  in Modern Browsers and Operating Systems}. In \bibinfo{booktitle}{\emph{Proc.
  of the 2015 10th ARES}}. \bibinfo{publisher}{IEEE Press},
  \bibinfo{pages}{174–179}.
\newblock
\showISBNx{9781467365901}


\bibitem[\protect\citeauthoryear{{Federal Networking Council}}{{Federal
  Networking Council}}{1997}]%
        {RFC-2146}
\bibfield{author}{\bibinfo{person}{{Federal Networking Council}}.}
  \bibinfo{year}{1997}\natexlab{}.
\newblock \bibinfo{booktitle}{\emph{{U.S. Government Internet Domain Names}}}.
\newblock \bibinfo{type}{RFC} 2146. \bibinfo{institution}{IETF}.
\newblock


\bibitem[\protect\citeauthoryear{Felt, Reeder, Ainslie, Harris, Walker,
  Thompson, Acer, Morant, and Consolvo}{Felt et~al\mbox{.}}{2019}]%
        {Felt2019}
\bibfield{author}{\bibinfo{person}{Adrienne~Porter Felt},
  \bibinfo{person}{Robert~W. Reeder}, \bibinfo{person}{Alex Ainslie},
  \bibinfo{person}{Helen Harris}, \bibinfo{person}{Max Walker},
  \bibinfo{person}{Christopher Thompson}, \bibinfo{person}{Mustafa~Emre Acer},
  \bibinfo{person}{Elisabeth Morant}, {and} \bibinfo{person}{Sunny Consolvo}.}
  \bibinfo{year}{2019}\natexlab{}.
\newblock \showarticletitle{{Rethinking connection security indicators}}. In
  \bibinfo{booktitle}{\emph{Proc. of 12th SOUPS}}. \bibinfo{publisher}{{USENIX}
  Association}, \bibinfo{pages}{1--14}.
\newblock
\showISBNx{9781931971317}


\bibitem[\protect\citeauthoryear{FEMA}{FEMA}{2004}]%
        {FEMA2004}
\bibfield{author}{\bibinfo{person}{FEMA}.} \bibinfo{year}{2004}\natexlab{}.
\newblock \bibinfo{title}{{Are you ready? An In-depth Guide to Citizen
  Preparedness}}.
\newblock
\newblock


\bibitem[\protect\citeauthoryear{FEMA}{FEMA}{2020a}]%
        {fema-ipaws20}
\bibfield{author}{\bibinfo{person}{FEMA}.} \bibinfo{year}{2020}\natexlab{a}.
\newblock \bibinfo{title}{Integrated Public Alert \& Warning System}.
\newblock
\newblock
\urldef\tempurl%
\url{https://www.fema.gov/integrated-public-alert-warning-system}
\showURL{%
\tempurl}


\bibitem[\protect\citeauthoryear{FEMA}{FEMA}{2020b}]%
        {fema-aa20}
\bibfield{author}{\bibinfo{person}{FEMA}.} \bibinfo{year}{2020}\natexlab{b}.
\newblock \bibinfo{title}{Organizations with Alerting Authority Complete and In
  Process}.
\newblock
\newblock
\urldef\tempurl%
\url{https://www.fema.gov/media-library/assets/documents/117152}
\showURL{%
\tempurl}


\bibitem[\protect\citeauthoryear{Fischer}{Fischer}{2020}]%
        {axios-trust20}
\bibfield{author}{\bibinfo{person}{Sara Fischer}.}
  \bibinfo{year}{2020}\natexlab{}.
\newblock \bibinfo{title}{Media wrestles with public trust as coronavirus
  intensifies}.
\newblock
\newblock
\urldef\tempurl%
\url{https://www.axios.com/media-public-trust-coronavirus-da69dd7f-4b8a-4793-ac52-dc1ed2c3e35f.html}
\showURL{%
\tempurl}


\bibitem[\protect\citeauthoryear{for Good}{for Good}{2018}]%
        {globaltrend-18}
\bibfield{author}{\bibinfo{person}{Nonprofit~Tech for Good}.}
  \bibinfo{year}{2018}\natexlab{}.
\newblock \bibinfo{title}{2018 Global Trends in Giving Report}.
\newblock
\newblock
\urldef\tempurl%
\url{https://givingreport.ngo/}
\showURL{%
\tempurl}


\bibitem[\protect\citeauthoryear{Forum}{Forum}{2017a}]%
        {cab-ballot185}
\bibfield{author}{\bibinfo{person}{CA/Browser Forum}.}
  \bibinfo{year}{2017}\natexlab{a}.
\newblock \bibinfo{title}{Ballot 185 -- Limiting the Lifetime of Certificates}.
\newblock
\newblock
\urldef\tempurl%
\url{https://cabforum.org/2017/02/24/ballot-185-limiting-lifetime-certificates/}
\showURL{%
\tempurl}


\bibitem[\protect\citeauthoryear{Forum}{Forum}{2017b}]%
        {cab-ballot193}
\bibfield{author}{\bibinfo{person}{CA/Browser Forum}.}
  \bibinfo{year}{2017}\natexlab{b}.
\newblock \bibinfo{title}{Ballot 193 -- 825-day Certificate Lifetimes}.
\newblock
\newblock
\urldef\tempurl%
\url{https://cabforum.org/2017/03/17/ballot-193-825-day-certificate-lifetimes/}
\showURL{%
\tempurl}


\bibitem[\protect\citeauthoryear{Forum}{Forum}{2019}]%
        {cab-ballotsc22}
\bibfield{author}{\bibinfo{person}{CA/Browser Forum}.}
  \bibinfo{year}{2019}\natexlab{}.
\newblock \bibinfo{title}{Ballot SC22 – Reduce Certificate Lifetimes (v2)}.
\newblock
\newblock
\urldef\tempurl%
\url{https://cabforum.org/2019/09/10/ballot-sc22-reduce-certificate-lifetimes-v2/}
\showURL{%
\tempurl}


\bibitem[\protect\citeauthoryear{Geiger}{Geiger}{2019}]%
        {pew-us-news19}
\bibfield{author}{\bibinfo{person}{Abigail~W. Geiger}.}
  \bibinfo{year}{2019}\natexlab{}.
\newblock \bibinfo{title}{Key findings about the online news landscape in
  America}.
\newblock
\newblock
\urldef\tempurl%
\url{https://pewrsr.ch/34CNdu3}
\showURL{%
\tempurl}


\bibitem[\protect\citeauthoryear{{Hao Yang}, Osterweil, Massey, {Songwu Lu},
  and {Lixia Zhang}}{{Hao Yang} et~al\mbox{.}}{2011}]%
        {HaoYang2011}
\bibfield{author}{\bibinfo{person}{{Hao Yang}}, \bibinfo{person}{Eric
  Osterweil}, \bibinfo{person}{Dan Massey}, \bibinfo{person}{{Songwu Lu}},
  {and} \bibinfo{person}{{Lixia Zhang}}.} \bibinfo{year}{2011}\natexlab{}.
\newblock \showarticletitle{{Deploying Cryptography in Internet-Scale Systems:
  A Case Study on DNSSEC}}.
\newblock \bibinfo{journal}{\emph{IEEE Transactions on Dependable and Secure
  Computing}} \bibinfo{volume}{8}, \bibinfo{number}{5} (\bibinfo{date}{sep}
  \bibinfo{year}{2011}), \bibinfo{pages}{656--669}.
\newblock
\showISSN{1545-5971}


\bibitem[\protect\citeauthoryear{Heinl, Giehl, Wiedermann, Plaga, and
  Kargl}{Heinl et~al\mbox{.}}{2019}]%
        {Heinl2019}
\bibfield{author}{\bibinfo{person}{Michael~P. Heinl},
  \bibinfo{person}{Alexander Giehl}, \bibinfo{person}{Norbert Wiedermann},
  \bibinfo{person}{Sven Plaga}, {and} \bibinfo{person}{Frank Kargl}.}
  \bibinfo{year}{2019}\natexlab{}.
\newblock \showarticletitle{MERCAT: A Metric for the Evaluation and
  Reconsideration of Certificate Authority Trustworthiness}. In
  \bibinfo{booktitle}{\emph{Proc. of the 2019 ACM SIGSAC CCSW}}.
  \bibinfo{publisher}{ACM Press}, \bibinfo{address}{New York, NY, USA},
  \bibinfo{pages}{1–15}.
\newblock
\showISBNx{9781450368261}


\bibitem[\protect\citeauthoryear{Hofmann}{Hofmann}{2019}]%
        {firefox-ev-19}
\bibfield{author}{\bibinfo{person}{Johann Hofmann}.}
  \bibinfo{year}{2019}\natexlab{}.
\newblock \bibinfo{title}{Intent to Ship: Move Extended Validation Information
  out of the URL bar}.
\newblock
\newblock
\urldef\tempurl%
\url{https://groups.google.com/d/msg/firefox-dev/6wAg_PpnlY4/C_DCyZm9AQAJ}
\showURL{%
\tempurl}


\bibitem[\protect\citeauthoryear{Holz, Braun, Kammenhuber, and Carle}{Holz
  et~al\mbox{.}}{2011}]%
        {Holz2011}
\bibfield{author}{\bibinfo{person}{Ralph Holz}, \bibinfo{person}{Lothar Braun},
  \bibinfo{person}{Nils Kammenhuber}, {and} \bibinfo{person}{Georg Carle}.}
  \bibinfo{year}{2011}\natexlab{}.
\newblock \showarticletitle{{The SSL landscape -- A Thorough Analysis of the
  X.509 PKI Using Active and Passive Measurement}}. In
  \bibinfo{booktitle}{\emph{Proc. of the ACM IMC '11}}. \bibinfo{publisher}{ACM
  Press}, \bibinfo{address}{New York, NY, USA}, \bibinfo{pages}{427}.
\newblock
\showISBNx{9781450310130}


\bibitem[\protect\citeauthoryear{Hughes and Chauhan}{Hughes and
  Chauhan}{2015}]%
        {Hughes2015}
\bibfield{author}{\bibinfo{person}{Amanda~Lee Hughes} {and}
  \bibinfo{person}{Apoorva Chauhan}.} \bibinfo{year}{2015}\natexlab{}.
\newblock \showarticletitle{{Online media as a means to affect public trust in
  emergency responders}}.
\newblock \bibinfo{journal}{\emph{ISCRAM 2015 Conference Proceedings}}
  (\bibinfo{year}{2015}), \bibinfo{pages}{182--192}.
\newblock
\showISBNx{9788271177881}


\bibitem[\protect\citeauthoryear{ICANNWiki}{ICANNWiki}{2017a}]%
        {icann-cctld}
\bibfield{author}{\bibinfo{person}{ICANNWiki}.}
  \bibinfo{year}{2017}\natexlab{a}.
\newblock \bibinfo{title}{Country code TLD}.
\newblock
\newblock
\urldef\tempurl%
\url{https://icannwiki.org/ccTLD}
\showURL{%
\tempurl}


\bibitem[\protect\citeauthoryear{ICANNWiki}{ICANNWiki}{2017b}]%
        {icann-gtld}
\bibfield{author}{\bibinfo{person}{ICANNWiki}.}
  \bibinfo{year}{2017}\natexlab{b}.
\newblock \bibinfo{title}{Generic TLD}.
\newblock
\newblock
\urldef\tempurl%
\url{https://icannwiki.org/GTLD}
\showURL{%
\tempurl}


\bibitem[\protect\citeauthoryear{ICANNWiki}{ICANNWiki}{2017c}]%
        {icann-sld}
\bibfield{author}{\bibinfo{person}{ICANNWiki}.}
  \bibinfo{year}{2017}\natexlab{c}.
\newblock \bibinfo{title}{SLD}.
\newblock
\newblock
\urldef\tempurl%
\url{https://icannwiki.org/SLD}
\showURL{%
\tempurl}


\bibitem[\protect\citeauthoryear{ICANNWiki}{ICANNWiki}{2017d}]%
        {icann-stld}
\bibfield{author}{\bibinfo{person}{ICANNWiki}.}
  \bibinfo{year}{2017}\natexlab{d}.
\newblock \bibinfo{title}{STLD}.
\newblock
\newblock
\urldef\tempurl%
\url{https://icannwiki.org/STLD}
\showURL{%
\tempurl}


\bibitem[\protect\citeauthoryear{Khan, Huo, Li, and Kanich}{Khan
  et~al\mbox{.}}{2015}]%
        {khan2015every}
\bibfield{author}{\bibinfo{person}{Mohammad~Taha Khan}, \bibinfo{person}{Xiang
  Huo}, \bibinfo{person}{Zhou Li}, {and} \bibinfo{person}{Chris Kanich}.}
  \bibinfo{year}{2015}\natexlab{}.
\newblock \showarticletitle{Every second counts: Quantifying the negative
  externalities of cybercrime via typosquatting}. In
  \bibinfo{booktitle}{\emph{2015 IEEE Symposium on Security and Privacy}}.
  IEEE, \bibinfo{pages}{135--150}.
\newblock


\bibitem[\protect\citeauthoryear{Kintis, Miramirkhani, Lever, Chen,
  Romero-G\'{o}mez, Pitropakis, Nikiforakis, and Antonakakis}{Kintis
  et~al\mbox{.}}{2017}]%
        {Kintis2017}
\bibfield{author}{\bibinfo{person}{Panagiotis Kintis}, \bibinfo{person}{Najmeh
  Miramirkhani}, \bibinfo{person}{Charles Lever}, \bibinfo{person}{Yizheng
  Chen}, \bibinfo{person}{Rosa Romero-G\'{o}mez}, \bibinfo{person}{Nikolaos
  Pitropakis}, \bibinfo{person}{Nick Nikiforakis}, {and} \bibinfo{person}{Manos
  Antonakakis}.} \bibinfo{year}{2017}\natexlab{}.
\newblock \showarticletitle{Hiding in Plain Sight: A Longitudinal Study of
  Combosquatting Abuse}. In \bibinfo{booktitle}{\emph{Proc. of the 2017 ACM
  SIGSAC CCS}}. \bibinfo{publisher}{ACM Press}, \bibinfo{address}{New York, NY,
  USA}, \bibinfo{pages}{569–586}.
\newblock
\showISBNx{9781450349468}


\bibitem[\protect\citeauthoryear{Korczynski, Wullink, Tajalizadehkhoob, Moura,
  Noroozian, Bagley, and Hesselman}{Korczynski et~al\mbox{.}}{2018}]%
        {Korczynski2018}
\bibfield{author}{\bibinfo{person}{Maciej Korczynski}, \bibinfo{person}{Maarten
  Wullink}, \bibinfo{person}{Samaneh Tajalizadehkhoob},
  \bibinfo{person}{Giovane C.~M. Moura}, \bibinfo{person}{Arman Noroozian},
  \bibinfo{person}{Drew Bagley}, {and} \bibinfo{person}{Cristian Hesselman}.}
  \bibinfo{year}{2018}\natexlab{}.
\newblock \showarticletitle{Cybercrime After the Sunrise: A Statistical
  Analysis of DNS Abuse in New GTLDs}. In \bibinfo{booktitle}{\emph{Proc. of
  the 2018 ACM ASIACCS}}. \bibinfo{publisher}{ACM}, \bibinfo{address}{New York,
  NY, USA}, \bibinfo{pages}{609–623}.
\newblock
\showISBNx{9781450355766}


\bibitem[\protect\citeauthoryear{Krombholz, Mayer, Schmiedecker, and
  Weippl}{Krombholz et~al\mbox{.}}{2017}]%
        {Krombholz2017}
\bibfield{author}{\bibinfo{person}{Katharina Krombholz},
  \bibinfo{person}{Wilfried Mayer}, \bibinfo{person}{Martin Schmiedecker},
  {and} \bibinfo{person}{Edgar Weippl}.} \bibinfo{year}{2017}\natexlab{}.
\newblock \showarticletitle{{``I have no idea what I'm doing'' -- on the
  usability of deploying HTTPS}}. In \bibinfo{booktitle}{\emph{Proc. of the
  26th USENIX Security Symposium}}. \bibinfo{publisher}{{USENIX} Association},
  \bibinfo{pages}{1339--1356}.
\newblock
\showISBNx{9781931971409}


\bibitem[\protect\citeauthoryear{Laurie, Langley, and Kasper}{Laurie
  et~al\mbox{.}}{2013}]%
        {RFC-6962}
\bibfield{author}{\bibinfo{person}{B. Laurie}, \bibinfo{person}{A. Langley},
  {and} \bibinfo{person}{E. Kasper}.} \bibinfo{year}{2013}\natexlab{}.
\newblock \bibinfo{booktitle}{\emph{{Certificate Transparency}}}.
\newblock \bibinfo{type}{RFC} 6962. \bibinfo{institution}{IETF}.
\newblock


\bibitem[\protect\citeauthoryear{Lian, Rescorla, Shacham, and Savage}{Lian
  et~al\mbox{.}}{2013}]%
        {Lian2013}
\bibfield{author}{\bibinfo{person}{Wilson Lian}, \bibinfo{person}{Eric
  Rescorla}, \bibinfo{person}{Hovav Shacham}, {and} \bibinfo{person}{Stefan
  Savage}.} \bibinfo{year}{2013}\natexlab{}.
\newblock \showarticletitle{{Measuring the practical impact of DNSSEC
  deployment}}. In \bibinfo{booktitle}{\emph{Proc. of the 22nd USENIX Security
  Symposium}}. \bibinfo{publisher}{USENIX Association},
  \bibinfo{pages}{573--587}.
\newblock
\showISBNx{9781931971034}


\bibitem[\protect\citeauthoryear{Liu, Tome, Zhang, Choffnes, Levin, Maggs,
  Mislove, Schulman, and Wilson}{Liu et~al\mbox{.}}{2015}]%
        {Liu2015}
\bibfield{author}{\bibinfo{person}{Yabing Liu}, \bibinfo{person}{Will Tome},
  \bibinfo{person}{Liang Zhang}, \bibinfo{person}{David Choffnes},
  \bibinfo{person}{Dave Levin}, \bibinfo{person}{Bruce Maggs},
  \bibinfo{person}{Alan Mislove}, \bibinfo{person}{Aaron Schulman}, {and}
  \bibinfo{person}{Christo Wilson}.} \bibinfo{year}{2015}\natexlab{}.
\newblock \showarticletitle{An End-to-End Measurement of Certificate Revocation
  in the Web’s PKI}. In \bibinfo{booktitle}{\emph{Proc. of the ACM IMC '15}}.
  \bibinfo{publisher}{ACM Press}, \bibinfo{address}{New York, NY, USA},
  \bibinfo{pages}{183–196}.
\newblock
\showISBNx{9781450338486}


\bibitem[\protect\citeauthoryear{Longstaff and Yang}{Longstaff and
  Yang}{2008}]%
        {Longstaff2008}
\bibfield{author}{\bibinfo{person}{P~H Longstaff} {and}
  \bibinfo{person}{Sung-Un Yang}.} \bibinfo{year}{2008}\natexlab{}.
\newblock \showarticletitle{{Communication Management and Trust: Their Role in
  Building Resilience to ``Surprises'' Such As Natural Disasters, Pandemic Flu,
  and Terrorism}}.
\newblock \bibinfo{journal}{\emph{Ecology and Society}} \bibinfo{volume}{13},
  \bibinfo{number}{1} (\bibinfo{year}{2008}), \bibinfo{pages}{1--14}.
\newblock


\bibitem[\protect\citeauthoryear{Manoj and Baker}{Manoj and Baker}{2007}]%
        {Manoj2007}
\bibfield{author}{\bibinfo{person}{B.S. Manoj} {and}
  \bibinfo{person}{Alexandra~Hubenko Baker}.} \bibinfo{year}{2007}\natexlab{}.
\newblock \showarticletitle{{Communication challenges in emergency response}}.
\newblock \bibinfo{journal}{\emph{Commun. ACM}} \bibinfo{volume}{50},
  \bibinfo{number}{3} (\bibinfo{date}{March} \bibinfo{year}{2007}),
  \bibinfo{pages}{51--53}.
\newblock
\showISSN{00010782}


\bibitem[\protect\citeauthoryear{Mishari, {De Cristofaro}, Defrawy, and
  Tsudik}{Mishari et~al\mbox{.}}{2009}]%
        {Mishari2009}
\bibfield{author}{\bibinfo{person}{Mishari~Al Mishari},
  \bibinfo{person}{Emiliano {De Cristofaro}}, \bibinfo{person}{Karim~El
  Defrawy}, {and} \bibinfo{person}{Gene Tsudik}.}
  \bibinfo{year}{2009}\natexlab{}.
\newblock \showarticletitle{{Harvesting SSL Certificate Data to Identify
  Web-Fraud}}.
\newblock \bibinfo{journal}{\emph{International Journal of Network Security}}
  \bibinfo{volume}{14}, \bibinfo{number}{6} (\bibinfo{date}{September}
  \bibinfo{year}{2009}), \bibinfo{pages}{324--338}.
\newblock
\showISSN{1816353X}


\bibitem[\protect\citeauthoryear{Neustar}{Neustar}{2020}]%
        {dotus-faqs}
\bibfield{author}{\bibinfo{person}{Neustar}.} \bibinfo{year}{2020}\natexlab{}.
\newblock \bibinfo{title}{Frequently Asked Questions about the .US Domain}.
\newblock
\newblock
\urldef\tempurl%
\url{https://www.about.us/faqs}
\showURL{%
\tempurl}


\bibitem[\protect\citeauthoryear{{Neustar, Inc.}}{{Neustar, Inc.}}{[n.d.]}]%
        {uscompliance}
\bibfield{author}{\bibinfo{person}{{Neustar, Inc.}}}
  \bibinfo{year}{[n.d.]}\natexlab{}.
\newblock \bibinfo{title}{{.US Compliance Report}}.
\newblock
\newblock
\urldef\tempurl%
\url{https://ns-cdn.neustar.biz/creative_services/biz/neustar/www/resources/domain-names/us-locality-compliance-report.pdf}
\showURL{%
\tempurl}


\bibitem[\protect\citeauthoryear{NIST}{NIST}{2020}]%
        {dnssec-gov}
\bibfield{author}{\bibinfo{person}{NIST}.} \bibinfo{year}{2020}\natexlab{}.
\newblock \bibinfo{title}{Estimating USG IPv6 \& DNSSEC External Service
  Deployment Status}.
\newblock
\newblock
\urldef\tempurl%
\url{https://fedv6-deployment.antd.nist.gov/cgi-bin/generate-gov}
\showURL{%
\tempurl}


\bibitem[\protect\citeauthoryear{O'Brien}{O'Brien}{2019}]%
        {chrome-ev-19}
\bibfield{author}{\bibinfo{person}{Devon O'Brien}.}
  \bibinfo{year}{2019}\natexlab{}.
\newblock \bibinfo{title}{Upcoming Change to Chrome's Identity Indicators}.
\newblock
\newblock
\urldef\tempurl%
\url{https://groups.google.com/a/chromium.org/d/msg/security-dev/h1bTcoTpfeI/jUTk1z7VAAAJ}
\showURL{%
\tempurl}


\bibitem[\protect\citeauthoryear{of~Agriculture}{of~Agriculture}{2018}]%
        {wayback-asapfarm}
\bibfield{author}{\bibinfo{person}{Missouri~Department of Agriculture}.}
  \bibinfo{year}{2018}\natexlab{}.
\newblock \bibinfo{title}{Missouri Agricultural Stewardship Assurance Program}.
\newblock
\newblock
\urldef\tempurl%
\url{https://web.archive.org/web/20180107233724/https://asap.farm/}
\showURL{%
\tempurl}


\bibitem[\protect\citeauthoryear{{Office of E-Government and Information
  Technology}}{{Office of E-Government and Information Technology}}{2009}]%
        {Technology2009}
\bibfield{author}{\bibinfo{person}{{Office of E-Government and Information
  Technology}}.} \bibinfo{year}{2009}\natexlab{}.
\newblock \bibinfo{title}{{Securing the Federal Government's Domain Name System
  Infrastructure}}.
\newblock , \bibinfo{numpages}{3}~pages.
\newblock


\bibitem[\protect\citeauthoryear{Osterweil, McPherson, and Zhang}{Osterweil
  et~al\mbox{.}}{2014}]%
        {osterweil2014shape}
\bibfield{author}{\bibinfo{person}{Eric Osterweil}, \bibinfo{person}{Danny
  McPherson}, {and} \bibinfo{person}{Lixia Zhang}.}
  \bibinfo{year}{2014}\natexlab{}.
\newblock \showarticletitle{The shape and size of threats: Defining a networked
  system's attack surface}. In \bibinfo{booktitle}{\emph{2014 IEEE 22nd ICNP}}.
  IEEE, \bibinfo{pages}{636--641}.
\newblock


\bibitem[\protect\citeauthoryear{Osterweil, Ryan, Massey, and Zhang}{Osterweil
  et~al\mbox{.}}{2008}]%
        {Osterweil2008}
\bibfield{author}{\bibinfo{person}{Eric Osterweil}, \bibinfo{person}{Michael
  Ryan}, \bibinfo{person}{Dan Massey}, {and} \bibinfo{person}{Lixia Zhang}.}
  \bibinfo{year}{2008}\natexlab{}.
\newblock \showarticletitle{Quantifying the Operational Status of the DNSSEC
  Deployment}. In \bibinfo{booktitle}{\emph{Proc. of the ACM IMC '08}}.
  \bibinfo{publisher}{ACM Press}, \bibinfo{address}{New York, NY, USA},
  \bibinfo{pages}{231–242}.
\newblock
\showISBNx{9781605583341}


\bibitem[\protect\citeauthoryear{Palen, Anderson, Mark, Martin, Sicker, Palmer,
  and Grunwald}{Palen et~al\mbox{.}}{2010}]%
        {Palen2010}
\bibfield{author}{\bibinfo{person}{Leysia Palen}, \bibinfo{person}{Kenneth~M.
  Anderson}, \bibinfo{person}{Gloria Mark}, \bibinfo{person}{James Martin},
  \bibinfo{person}{Douglas Sicker}, \bibinfo{person}{Martha Palmer}, {and}
  \bibinfo{person}{Dirk Grunwald}.} \bibinfo{year}{2010}\natexlab{}.
\newblock \showarticletitle{A Vision for Technology-Mediated Support for Public
  Participation \& Assistance in Mass Emergencies \& Disasters}. In
  \bibinfo{booktitle}{\emph{Proc. of the 2010 ACM-BCS}}.
  \bibinfo{publisher}{BCS Learning \& Development Ltd.},
  \bibinfo{address}{Swindon, GBR}, Article \bibinfo{articleno}{8},
  \bibinfo{numpages}{12}~pages.
\newblock
\showISBNx{9781450301923}


\bibitem[\protect\citeauthoryear{Paton}{Paton}{2007}]%
        {Paton2007}
\bibfield{author}{\bibinfo{person}{Douglas Paton}.}
  \bibinfo{year}{2007}\natexlab{}.
\newblock \showarticletitle{{Preparing for natural hazards: The role of
  community trust}}.
\newblock \bibinfo{journal}{\emph{Disaster Prevention and Management}}
  \bibinfo{volume}{16}, \bibinfo{number}{3} (\bibinfo{year}{2007}),
  \bibinfo{pages}{370--379}.
\newblock
\showISBNx{0957409091095}
\showISSN{09653562}


\bibitem[\protect\citeauthoryear{Postel and Reynolds}{Postel and
  Reynolds}{1984}]%
        {RFC-920}
\bibfield{author}{\bibinfo{person}{J. Postel} {and} \bibinfo{person}{J.K.
  Reynolds}.} \bibinfo{year}{1984}\natexlab{}.
\newblock \bibinfo{booktitle}{\emph{{Domain requirements}}}.
\newblock \bibinfo{type}{RFC} 920. \bibinfo{institution}{IETF}.
\newblock


\bibitem[\protect\citeauthoryear{Project}{Project}{2020}]%
        {spamhaus20}
\bibfield{author}{\bibinfo{person}{Spamhaus Project}.}
  \bibinfo{year}{2020}\natexlab{}.
\newblock \bibinfo{title}{The World's Most Abused TLDs}.
\newblock
\newblock
\urldef\tempurl%
\url{https://www.spamhaus.org/statistics/tlds/}
\showURL{%
\tempurl}


\bibitem[\protect\citeauthoryear{Reporter}{Reporter}{2020}]%
        {trustpoll-20}
\bibfield{author}{\bibinfo{person}{Morning Consult/The~Hollywood Reporter}.}
  \bibinfo{year}{2020}\natexlab{}.
\newblock \bibinfo{title}{National Tracking Poll \#200342 -- Crosstabulation
  Results}.
\newblock
\newblock
\urldef\tempurl%
\url{https://morningconsult.com/wp-content/uploads/2020/03/200342_crosstabs_HOLLYWOOD_Adults_v2_JB-1.pdf}
\showURL{%
\tempurl}


\bibitem[\protect\citeauthoryear{Research}{Research}{2020}]%
        {dnssec-tld}
\bibfield{author}{\bibinfo{person}{ICANN Research}.}
  \bibinfo{year}{2020}\natexlab{}.
\newblock \bibinfo{title}{TLD DNSSEC Report}.
\newblock
\newblock
\urldef\tempurl%
\url{http://stats.research.icann.org/dns/tld_report/}
\showURL{%
\tempurl}


\bibitem[\protect\citeauthoryear{Roberts, Goldschlag, Walter, Chung, Mislove,
  and Levin}{Roberts et~al\mbox{.}}{2019}]%
        {Roberts2019}
\bibfield{author}{\bibinfo{person}{Richard Roberts}, \bibinfo{person}{Yaelle
  Goldschlag}, \bibinfo{person}{Rachel Walter}, \bibinfo{person}{Taejoong
  Chung}, \bibinfo{person}{Alan Mislove}, {and} \bibinfo{person}{Dave Levin}.}
  \bibinfo{year}{2019}\natexlab{}.
\newblock \showarticletitle{You Are Who You Appear to Be: A Longitudinal Study
  of Domain Impersonation in TLS Certificates}. In
  \bibinfo{booktitle}{\emph{Proc. of the 2019 ACM SIGSAC CCS}}.
  \bibinfo{publisher}{ACM Press}, \bibinfo{address}{New York, NY, USA},
  \bibinfo{pages}{2489–2504}.
\newblock
\showISBNx{9781450367479}


\bibitem[\protect\citeauthoryear{Roberts, Lulli, Raut, Fulton, and
  Levin}{Roberts et~al\mbox{.}}{2020}]%
        {roberts20}
\bibfield{author}{\bibinfo{person}{Richard Roberts}, \bibinfo{person}{Daniela
  Lulli}, \bibinfo{person}{Abolee Raut}, \bibinfo{person}{Kelsey Fulton}, {and}
  \bibinfo{person}{Dave~Levin Levin}.} \bibinfo{year}{2020}\natexlab{}.
\newblock \showarticletitle{Mental Models of Domain Names and URLs}. In
  \bibinfo{booktitle}{\emph{Proc. of SOUPS}}. \bibinfo{publisher}{{USENIX}
  Association}, \bibinfo{pages}{5}.
\newblock


\bibitem[\protect\citeauthoryear{Rose}{Rose}{2012}]%
        {Rose2012}
\bibfield{author}{\bibinfo{person}{Scott Rose}.}
  \bibinfo{year}{2012}\natexlab{}.
\newblock \showarticletitle{Progress of DNS Security Deployment in the Federal
  Government}. In \bibinfo{booktitle}{\emph{Proc. of the 26th LISA}}.
  \bibinfo{publisher}{USENIX Association}, \bibinfo{pages}{223–228}.
\newblock


\bibitem[\protect\citeauthoryear{Santesson, Myers, Ankney, Malpani, Galperin,
  and Adams}{Santesson et~al\mbox{.}}{2013}]%
        {RFC-6960}
\bibfield{author}{\bibinfo{person}{S. Santesson}, \bibinfo{person}{M. Myers},
  \bibinfo{person}{R. Ankney}, \bibinfo{person}{A. Malpani},
  \bibinfo{person}{S. Galperin}, {and} \bibinfo{person}{C. Adams}.}
  \bibinfo{year}{2013}\natexlab{}.
\newblock \bibinfo{booktitle}{\emph{{X.509 Internet Public Key Infrastructure
  Online Certificate Status Protocol - OCSP}}}.
\newblock \bibinfo{type}{RFC} 6960. \bibinfo{institution}{IETF}.
\newblock


\bibitem[\protect\citeauthoryear{Scheitle, Gasser, Nolte, Amann, Brent, Carle,
  Holz, Schmidt, and W{\"{a}}hlisch}{Scheitle et~al\mbox{.}}{2018}]%
        {Scheitle2018}
\bibfield{author}{\bibinfo{person}{Quirin Scheitle}, \bibinfo{person}{Oliver
  Gasser}, \bibinfo{person}{Theodor Nolte}, \bibinfo{person}{Johanna Amann},
  \bibinfo{person}{Lexi Brent}, \bibinfo{person}{Georg Carle},
  \bibinfo{person}{Ralph Holz}, \bibinfo{person}{Thomas~C. Schmidt}, {and}
  \bibinfo{person}{Matthias W{\"{a}}hlisch}.} \bibinfo{year}{2018}\natexlab{}.
\newblock \showarticletitle{{The Rise of Certificate Transparency and Its
  Implications on the Internet Ecosystem}}. In \bibinfo{booktitle}{\emph{Proc.
  of the ACM IMC '18}}. \bibinfo{publisher}{ACM Press}, \bibinfo{address}{New
  York, NY, USA}, \bibinfo{pages}{343--349}.
\newblock
\showISBNx{9781450356190}


\bibitem[\protect\citeauthoryear{{Schwittmann}, {Wander}, and
  {Weis}}{{Schwittmann} et~al\mbox{.}}{2019}]%
        {Schwittmann2019}
\bibfield{author}{\bibinfo{person}{L. {Schwittmann}}, \bibinfo{person}{M.
  {Wander}}, {and} \bibinfo{person}{T. {Weis}}.}
  \bibinfo{year}{2019}\natexlab{}.
\newblock \showarticletitle{Domain Impersonation is Feasible: A Study of CA
  Domain Validation Vulnerabilities}. In \bibinfo{booktitle}{\emph{2019 IEEE
  European Symposium on Security and Privacy (EuroS\&P)}}.
  \bibinfo{publisher}{IEEE Press}, \bibinfo{pages}{544--559}.
\newblock


\bibitem[\protect\citeauthoryear{Seckler, Heinz, Forde, Tuch, and
  Opwis}{Seckler et~al\mbox{.}}{2015}]%
        {Seckler2015}
\bibfield{author}{\bibinfo{person}{Mirjam Seckler}, \bibinfo{person}{Silvia
  Heinz}, \bibinfo{person}{Seamus Forde}, \bibinfo{person}{Alexandre~N. Tuch},
  {and} \bibinfo{person}{Klaus Opwis}.} \bibinfo{year}{2015}\natexlab{}.
\newblock \showarticletitle{{Trust and distrust on the web: User experiences
  and website characteristics}}.
\newblock \bibinfo{journal}{\emph{Computers in Human Behavior}}
  \bibinfo{volume}{45} (\bibinfo{year}{2015}), \bibinfo{pages}{39--50}.
\newblock
\showISSN{07475632}


\bibitem[\protect\citeauthoryear{Shu and Schieber}{Shu and Schieber}{2020}]%
        {techcrunch-joint20}
\bibfield{author}{\bibinfo{person}{Catherine Shu} {and}
  \bibinfo{person}{Jonathan Schieber}.} \bibinfo{year}{2020}\natexlab{}.
\newblock \bibinfo{title}{Facebook, Reddit, Google, LinkedIn, Microsoft,
  Twitter and YouTube issue joint statement on misinformation}.
\newblock
\newblock
\urldef\tempurl%
\url{https://tcrn.ch/2xJXrg8}
\showURL{%
\tempurl}


\bibitem[\protect\citeauthoryear{Singanamalla, Jang, Anderson, Kohno, and
  Heimerl}{Singanamalla et~al\mbox{.}}{2020}]%
        {10.1145/3419394.3423645}
\bibfield{author}{\bibinfo{person}{Sudheesh Singanamalla},
  \bibinfo{person}{Esther Han~Beol Jang}, \bibinfo{person}{Richard Anderson},
  \bibinfo{person}{Tadayoshi Kohno}, {and} \bibinfo{person}{Kurtis Heimerl}.}
  \bibinfo{year}{2020}\natexlab{}.
\newblock \showarticletitle{Accept the Risk and Continue: Measuring the Long
  Tail of Government https Adoption}. In \bibinfo{booktitle}{\emph{Proc. of the
  ACM IMC '20}}. \bibinfo{publisher}{ACM}, \bibinfo{address}{New York, NY,
  USA}, \bibinfo{pages}{577–597}.
\newblock
\showISBNx{9781450381383}


\bibitem[\protect\citeauthoryear{{Sophos Labs}}{{Sophos Labs}}{2020}]%
        {nullSL2020}
\bibfield{author}{\bibinfo{person}{{Sophos Labs}}.}
  \bibinfo{year}{2020}\natexlab{}.
\newblock \bibinfo{title}{Facing down the myriad threats tied to {COVID}-19}.
\newblock
\newblock
\urldef\tempurl%
\url{https://news.sophos.com/en-us/2020/04/14/covidmalware/}
\showURL{%
\tempurl}


\bibitem[\protect\citeauthoryear{Suzuki, Chiba, Yoneya, Mori, and Goto}{Suzuki
  et~al\mbox{.}}{2019}]%
        {Suzuki2019}
\bibfield{author}{\bibinfo{person}{Hiroaki Suzuki}, \bibinfo{person}{Daiki
  Chiba}, \bibinfo{person}{Yoshiro Yoneya}, \bibinfo{person}{Tatsuya Mori},
  {and} \bibinfo{person}{Shigeki Goto}.} \bibinfo{year}{2019}\natexlab{}.
\newblock \showarticletitle{ShamFinder: An Automated Framework for Detecting
  IDN Homographs}. In \bibinfo{booktitle}{\emph{Proc. of the ACM IMC '19}}.
  \bibinfo{publisher}{ACM Press}, \bibinfo{address}{New York, NY, USA},
  \bibinfo{pages}{449–462}.
\newblock
\showISBNx{9781450369480}


\bibitem[\protect\citeauthoryear{Tian, Jan, Hu, Yao, and Wang}{Tian
  et~al\mbox{.}}{2018}]%
        {Tian2018}
\bibfield{author}{\bibinfo{person}{Ke Tian}, \bibinfo{person}{Steve T.~K. Jan},
  \bibinfo{person}{Hang Hu}, \bibinfo{person}{Danfeng Yao}, {and}
  \bibinfo{person}{Gang Wang}.} \bibinfo{year}{2018}\natexlab{}.
\newblock \showarticletitle{{Needle in a Haystack: Tracking Down Elite Phishing
  Domains in the Wild}}. In \bibinfo{booktitle}{\emph{Proc. of the ACM IMC
  '18}}. \bibinfo{publisher}{ACM Press}, \bibinfo{address}{New York, NY, USA},
  \bibinfo{pages}{429--442}.
\newblock
\showISBNx{9781450356190}


\bibitem[\protect\citeauthoryear{Tursunbayeva, Franco, and
  Pagliari}{Tursunbayeva et~al\mbox{.}}{2017}]%
        {Tursunbayeva2017}
\bibfield{author}{\bibinfo{person}{Aizhan Tursunbayeva},
  \bibinfo{person}{Massimo Franco}, {and} \bibinfo{person}{Claudia Pagliari}.}
  \bibinfo{year}{2017}\natexlab{}.
\newblock \showarticletitle{{Use of social media for e-Government in the public
  health sector: A systematic review of published studies}}.
\newblock \bibinfo{journal}{\emph{Government Information Quarterly}}
  \bibinfo{volume}{34}, \bibinfo{number}{2} (\bibinfo{date}{apr}
  \bibinfo{year}{2017}), \bibinfo{pages}{270--282}.
\newblock
\showISSN{0740624X}


\bibitem[\protect\citeauthoryear{Walther, Wang, and Loh}{Walther
  et~al\mbox{.}}{2004}]%
        {Walther2004}
\bibfield{author}{\bibinfo{person}{Joseph~B. Walther}, \bibinfo{person}{Zuoming
  Wang}, {and} \bibinfo{person}{Tracy Loh}.} \bibinfo{year}{2004}\natexlab{}.
\newblock \showarticletitle{{The effect of top-level domains and advertisements
  on health web-site credibility}}.
\newblock \bibinfo{journal}{\emph{Journal of Medical Internet Research}}
  \bibinfo{volume}{6}, \bibinfo{number}{3} (\bibinfo{year}{2004}),
  \bibinfo{pages}{1--11}.
\newblock
\showISSN{14388871}


\bibitem[\protect\citeauthoryear{Wander and Weis}{Wander and Weis}{2013}]%
        {Wander2013}
\bibfield{author}{\bibinfo{person}{Matth{\"{a}}us Wander} {and}
  \bibinfo{person}{Torben Weis}.} \bibinfo{year}{2013}\natexlab{}.
\newblock \showarticletitle{{Measuring Occurrence of DNSSEC Validation}}. In
  \bibinfo{booktitle}{\emph{Passive and Active Measurement}},
  \bibfield{editor}{\bibinfo{person}{Matthew Roughan} {and}
  \bibinfo{person}{Rocky Chang}} (Eds.). \bibinfo{publisher}{Springer Berlin
  Heidelberg}, \bibinfo{address}{Berlin, Heidelberg},
  \bibinfo{pages}{125--134}.
\newblock
\showISBNx{978-3-642-36516-4}


\bibitem[\protect\citeauthoryear{Wang}{Wang}{2015}]%
        {wang2015revisit}
\bibfield{author}{\bibinfo{person}{Zheng Wang}.}
  \bibinfo{year}{2015}\natexlab{}.
\newblock \showarticletitle{A revisit of DNS Kaminsky cache poisoning attacks}.
  In \bibinfo{booktitle}{\emph{2015 IEEE GLOBECOM}}. IEEE,
  \bibinfo{pages}{1--6}.
\newblock


\bibitem[\protect\citeauthoryear{WHO}{WHO}{2020}]%
        {who-infodemic20}
\bibfield{author}{\bibinfo{person}{WHO}.} \bibinfo{year}{2020}\natexlab{}.
\newblock \bibinfo{title}{Novel Coronavirus (2019-nCoV) Situation Report - 13}.
\newblock
\newblock
\urldef\tempurl%
\url{https://apps.who.int/iris/handle/10665/330778}
\showURL{%
\tempurl}


\end{thebibliography}

\appendix
\section*{Appendix}
\label{sec:app}

\setcounter{table}{0}
\renewcommand\thetable{A.\Roman{table}}
\begin{table}[!h]
	\scriptsize
	\caption{Regular expressions applied on an AA name to categorize its field of operation (in order of application).}
	\label{tab:domain-classes}
	\vspace{-9pt}
	\begin{tabularx}{\columnwidth}{l >{\ttfamily}X}
		\toprule
		Category & \textnormal{Regular expression} \\
		\midrule
		Military & [\textasciicircum[:alnum:]]fort|\textasciicircum fort|army|missile|base|pfpa\\\dhline
		Governmental & county|counties|city|commission|borough|town|village|parish| authority|council| government|national|aviation|correction\\\dhline
		Educational & university\\\dhline
		Law Enforcement & police|sheriff|investigation|patrol|intelligence|'homeland security'|'law enforcement'\\\dhline
		Public Safety & 911|'9-1-1'|emergency|ema|eom|ohsep|fire|safety| communication|dispatch\\
		\bottomrule
	\end{tabularx}
	\vspace{-.6cm}
\end{table}

 \begin{table}[!h]
	\setlength{\tabcolsep}{1.5pt}
	\scriptsize
	\caption{Count of unique hosts with at least one publicly logged certificate per issuer for popular CAs. The last row shows the sum of unique host for all observed CAs.}
	\vspace{-9pt}
	\label{tab:total-certs-per-ca}
	\begin{minipage}{\columnwidth}
		\begin{tabularx}{\columnwidth}{X c c c c c c c c c c c}
			\toprule
			& \multicolumn{11}{c}{Year}
			\\
			\cmidrule{2-12}
			CA \hspace{2cm} & '09 & '10 & '11 & '12 & '13 & '14 & '15 & '16 & '17 & '18 & '19\\
			\midrule
			Comodo\textsuperscript{$\dagger$} &    3 &    7 &   15 &   21 &   29 &   38 &   62 &   92 &  238 &  304 &  299\\\dhline
			DigiCert &   31 &   53 &   70 &   83 &   92 &  105 &  120 &  133 &  146 &  263 &  281\\\dhline
			Entrust &    7 &   13 &   22 &   25 &   34 &   32 &   33 &   39 &   40 &   44 &   48\\\dhline
			GeoTrust\textsuperscript{$\ddagger$} &    0 &    5 &   29 &   49 &   54 &   59 &   63 &   67 &   68 &   61 &   29\\\dhline
			GoDaddy &   25 &   54 &   80 &  109 &  141 &  183 &  215 &  249 &  290 &  330 &  347\\\dhline
			LetsEncrypt\textsuperscript{$\dagger\dagger$} &    0 &    0 &    0 &    0 &    0 &    0 &    0 &   29 &  102 &  210 &  335\\\dhline
			Sectigo &    0 &    0 &    0 &    0 &    0 &    0 &    0 &    0 &    0 &    0 &  228\\\dhline
			Verisign\textsuperscript{$\ddagger$} &   18 &   45 &   49 &   43 &   43 &   42 &   35 &   27 &   17 &    6 &    0\\\midrule
			All observed CAs &  122 &  244 &  298 &  356 &  398 &  458 &  517 &  630 &  830 & 1012 & 1109\\
			
			\bottomrule
		\end{tabularx}
		\smallskip
		
		\scriptsize{\textsuperscript{$\dagger$}~Rebranded to Sectigo in 2018.} \quad
		\scriptsize{\textsuperscript{$\ddagger$}~Acquired by DigiCert in 2017.} \quad
		\scriptsize{\textsuperscript{$\dagger\dagger$}~Beta in 2015; public in~2016.}
	\end{minipage}
\end{table}

\label{lastpage}

\end{document}